# L1 and off Sun-Earth line visible-light imaging of Earth-directed CMEs: An analysis of inconsistent observations

**Richard A. Harrison[1], Jackie A. Davies[1], David Barnes[1] and Christian Möstl[2]**

[1]RAL Space, STFC Rutherford Appleton Laboratory, Harwell Campus, Didcot OX11 0QX, UK

[2]Austrian Space Weather Office, GeoSphere Austria, 8020 Graz, Austria

Corresponding author: first and last name (richard.harrison@stfc.ac.uk)

[Orcid IDs: R.A. Harrison 0000-0002-0843-8045; J.A. Davies 0000-0001-9865-9281; D. Barnes 0000-0003-1137-8220; C. Mostl 0000-0001-6868-4152]

## Key Points:

- The identification of Earth-directed coronal mass ejections (CMEs) is compared for observations from L1 and off the Sun-Earth line, for 2011.
- We show a number of CMEs identified as Earth directed, with consistent in-situ ICME data, with no identification from L1 data.
- Given the number of ICMEs recorded at L1 in 2011, the number of inconsistent events is of concern.

## Abstract

The efficacy of coronal mass ejection (CME) observations as a key input to space weather forecasting is explored by comparing on and off Sun-Earth line observations from the ESA/NASA SOHO and NASA STEREO spacecraft. A comparison is made of CME catalogues based on L1 coronagraph imagery and off Sun-Earth line coronagraph and heliospheric imager (HI) observations, for the year 2011. Analysis reveals inconsistencies in the identification of a number of potentially Earth-directed CMEs. The catalogues reflect our ability to identify and characterise CMEs, so any discrepancies can impact our prediction of Earth-directed CMEs. We show that 15 CMEs, which were observed by STEREO, that had estimated directions compatible with Earth-directed events, had no identified halo/partial halo counterpart listed in the L1 coronagraph CME catalogue. In-situ data confirms that for 9 of these there was a consistent L1 Interplanetary CME (ICME). The number of such 'discrepant' events is significant compared to the number of ICMEs recorded at L1 in 2011, stressing the need to address space weather monitoring capabilities, particularly with the inclusion of off Sun-Earth line observation. While the study provides evidence that some halo CMEs are simply not visible in near-Earth coronagraph imagery, there is evidence that some halo CMEs viewed from L1 are compromised by preceding CME remnants or the presence of multiple-CMEs. This underlines (1) the value of multiple vantage point CME observation, and (2) the benefit of off Sun-Earth line platform heliospheric imaging, and coronagraph imaging, for the efficient identification and tracking of Earth-directed events.

## Plain Language Summary

Coronal Mass Ejections (CMEs) can impact human activities and technological assets and are continuously monitored by dedicated space weather operations centres. This study identifies inconsistencies between the identification of potentially Earth-directed CMEs in observations made from a near Earth vantage point and those made from off the Sun-Earth line. Almost a third of a set of potentially Earth-directed CMEs that were identified in an off Sun-Earth line heliospheric imager catalogue from 2011, which were subsequently evidenced as Interplanetary CMEs (ICMEs) in in-situ data near Earth, were not catalogued as halo or partial halo events in near Earth coronagraph imagery. Given the number of such discrepant events, compared to the total number of ICMEs recorded near Earth in 2011, it is important to understand the inconsistencies and to ensure that future space weather monitoring strategies cater for them.

## 1 Introduction

At the heart of space weather forecasting is the identification and tracking of coronal mass ejections (CMEs) in visible light images. The aim of this paper is to compare the use of such imaging, both from instruments stationed on and off the Sun-Earth line in the identification and characterisation of Earth-impacting CMEs. In particular, we undertake this analysis through comparison of established CME catalogues, and in so doing rely on the greater experience of those that routinely generate the catalogues rather than performing such identification/characterisation ourselves. This is important because the existing catalogues illustrate our ability to accurately identify and characterise CMEs from different platforms and locations, although we note, of course, that CME catalogues for different space-borne instruments are compiled using varying methods and analysis techniques. The visible-light CME catalogues that we have chosen to compare are those that are endorsed by the relevant instrument teams and are widely used by the research community. They are catalogues of CMEs observed using the Heliospheric Imager (HI) instruments (Eyles et al., 2009) and outer coronagraphs (COR-2; Howard et al., 2008) aboard the NASA STEREO (Solar Terrestrial Relations Observatory; Kaiser et al., 2008) spacecraft and using the Large Angle

Spectrometric Coronagraph (LASCO; Brueckner et al., 1995) aboard the ESA/NASA SOHO (Solar and Heliospheric Observatory; Domingo et al., 1995) spacecraft.

The STEREO mission was launched in October 2006, with science operations commencing in April 2007, with a primary goal of studying CMEs from launch to 1 AU and beyond, and in particular those heading towards Earth. Thus, we have had STEREO observations of CMEs in the inner heliosphere for over 15 years. The twin STEREO spacecraft were inserted into near-1AU heliocentric orbits, with one spacecraft (STEREO-A) orbiting ahead of the Earth and the other spacecraft (STEREO-B) orbiting behind the Earth; both spacecraft move relative to the Sun-Earth line by about 22.5$^o$ per year. Thus, STEREO provides the ability to observe CMEs from a vantage point off the Sun-Earth line. Contact with the STEREO-B spacecraft was lost in 2014, but STEREO-A is still fully operational.

Each STEREO spacecraft carries a SECCHI (Sun Earth Connection Coronal and Heliospheric Investigation; Howard et al., 2008) instrument package comprising an EUV imager, an inner and outer coronagraph (COR-1 and COR-2 respectively) and an HI. Here, we focus on HI and COR-2. Each STEREO HI instrument includes two visible-light cameras, known as HI-1 and HI-2. We use the nomenclature HI-1A, HI-1B, HI-2A and HI-2B to denote HI-1 on STEREO-A, HI-1 on STEREO-B, HI-2 on STEREO-A and HI-2 on STEREO-B, respectively (Eyles et al., 2009; Harrison et al., 2008). We also use the notation HI-A and HI-B to denote the HI cameras on STEREO -A and STEREO-B, respectively. In nominal operations, the fields of view of the HI-1 and HI-2 cameras are both centred on the ecliptic plane, with boresights offset from Sun-centre by 14$^o$ and 53.7$^o$ elongation, respectively. With fields of view of 20$^o$ and 70$^o$ diameter, the HI-1 and HI-2 cameras allow the imaging and tracking of CMEs, between them, from elongations of 4$^o$ to 88.7$^o$ near the ecliptic. The COR-2 instrument on each STEREO spacecraft has a field of view extending from 2.5 to 15 solar radii from Sun-centre, with complete 360$^o$ position angle (PA) coverage around the solar disc.

The SOHO spacecraft was launched in 1995 and is stationed in an Earth-Sun L1 orbit. The SOHO LASCO instrument is comprised of a set of three coronagraphs, known as C1, C2 and C3. Since 1997, only the C2 and C3 visible-light coronagraphs have been operational, with a combined field of view extending from 1.5 to 30 solar radii from Sun-centre (C2 from 2 to 6 solar radii and C3 from 3.7 to 30 solar radii, i.e. out to 7.5$^o$ elongation), each with near-full 360$^o$ PA coverage.

The original rationale for the current study was to compare L1 CME images to those made by STEREO when the STEREO spacecraft were near to the Lagrange L4 and L5 points (60$^o$ ahead and 60$^o$ behind the Earth, respectively). This recognises, in particular, that the L5 point is considered to be optimal for space weather applications. For example, the ESA Vigil (formerly, Lagrange) space weather mission (Gibney, 2017), if adopted, will be stationed at L5. Such an off-Sun-Earth line position clearly provides an excellent vantage point for observing Earth-directed CMEs, which, from near Earth, would be seen as halo events in coronagraph images as the CMEs expand from behind the instrument occulter, resulting in ambiguity between CME angular width and speed (Millward et al., 2013). Hence, ideally, we would examine the STEREO data for the period from late 2008 to early autumn 2010, as the timeframe where the STEREO spacecraft were 60$^o$ +/- 10$^o$ behind (STEREO-B) and in front (STEREO-A) of the Earth in its orbit. However, this coincided with the deep solar minimum of 2008-10, making it a very poor timeframe to study CMEs at all, let alone CME arrivals at Earth. As a compromise, we selected the year 2011 for our study because of the increase in solar activity. Between January and December 2011 STEREO-A drifted from 85.6$^o$ to 107.1$^o$ west of the Sun-Earth line and STEREO-B drifted from 90.0$^o$ to 110.9$^o$ east of the Sun-Earth line. These locations are beyond L4 and L5 from the perspective of Earth but do enable adequate off Sun-Earth line observations of Earth-directed CMEs.

With the return of STEREO-A to within 20$^{o}$ of L5 from late 2019 to late 2021, there is another potential opportunity to repeat the current study. However, this, again, corresponds to a time of low solar activity and STEREO-B was no longer operational.

## 2 The Catalogues

CMEs imaged by STEREO/HI have been catalogued in the context of the HELCATS (Heliospheric Cataloguing, Analysis and Techniques Service) project (www.helcats-fp7.eu), which was devised to construct and disseminate catalogues of CMEs and CIRs/SIRs (Co-rotating/Stream Interaction Regions) based on a variety of modelling techniques, and to assess the efficacy of those techniques through comparison with ground truth measurements. HELCATS HICAT (HI CATalogue; Harrison et al., 2018) is a catalogue of CMEs detected visually in STEREO/HI imagery, and is still maintained as part of the HELCATS catalogue product-set (see 'products' tab on the HELCATS website), despite the fact that the HELCATS project formally ended in 2017. HICAT tabulates the CME observational parameters, including the date and time of the first observation of the CME in the HI-1 field of view, an estimation of its northern and southern PA extent, and a measure of the CME observation quality.

As first demonstrated by Rouillard et al. (2008) and Sheeley et al. (2008), it is possible to analyse the time-elongation profile of a solar wind feature observed from a single STEREO HI instrument to estimate its radial speed, launch time and 3D propagation direction in the plane defined by the PA in which the feature is tracked, based on geometric considerations of wide-angle imaging. Their initial analysis was performed for blobs entrained in co-rotating interaction regions, but the effect has been exploited on many occasions for the analysis of CMEs (e.g. Harrison et al., 2012; Mishra et al., 2015; Möstl et al., 2014).

Based on such simple geometric modelling, an augmented HELCATS catalogue, HIGeoCAT (Barnes et al.., 2019), was produced, which also includes kinematic parameters for the majority of the HICAT CMEs, i.e. radial speed of the CME apex, the estimated CME launch time and estimated propagation direction of the CME apex in HEE (Heliocentric Earth Equatorial) coordinates (which correspond to an estimated source longitude and latitude if constant direction from onset is assumed), derived from the application of several single-spacecraft geometrical modelling techniques. These parameters are generated via multi-variant fitting of the time-elongation profile along a fixed position angle derived from tracking the CME through a time-elongation map (J-map) derived from STEREO HI-1 and HI-2 data. The techniques are based on the assumption that the CMEs adopt simple geometrical forms that propagate radially from the Sun at a constant speed, expanding self-similarly as they propagate. The geometrical models used to generate HIGeoCAT are (1) the Fixed-Phi Fitting (FPF) technique (Kahler and Webb, 2007), in which the CME is assumed to adopt a point-like morphology, (2) the Harmonic Mean Fitting (HMF) technique (Lugaz et al., 2009), which models the CME as an expanding circle anchored at Sun-centre, and (3) the Self Similar Expansion Fitting (SSEF) technique (Davies et al., 2012), in which the CME is modelled as an outward propagating self-similarly expanding circle with any fixed half-width between 0$^{o}$ (corresponding to FPF) and 90$^{o}$ (corresponding to HMF; 30$^{o}$ is used in the case of HELCATS, subsequently referred to as SSEF30 (Barnes et al., 2019).

HICAT CMEs are identified as 'poor', 'fair' or 'good', based on the observational clarity of each event. This classification is obviously somewhat subjective, and approximately 20% of the CMEs in HICAT are classed as poor. These events are excluded from HIGeoCAT.

The HIGeoCAT catalogue includes 116 and 122 CMEs for 2011 for STEREO-A and STEREO-B, respectively (compared with 159 and 148 in HICAT for that year). The two spacecraft were separated by an angle (spacecraft-Sun-Spacecraft angle, centred approximately on the Sun-Earth line) of between 175$^o$ to 218$^o$ during this period, and the majority of the STEREO-A and STEREO-B entries were imaged by both HI instruments. Indeed, a visual inspection of the 116 CMEs, as reported in the HELCATS HIJoinCAT catalogue (Barnes et al., 2020), which lists the HI-A and HI-B CMEs that are considered to be common, suggests that 98 were recorded in both the HI-A and HI-B HIGeoCAT entries.

The number of HIGeoCAT CMEs differs between catalogued HI-A and HI-B entries; we do not see a one to one correlation between the instruments. The good, fair and poor classifications of HICAT are somewhat subjective. So, for example, an event that we classify as fair in HI-A might be classed as poor in HI-B (or vice versa); the latter would be excluded from HIGeoCAT. This might relate to observational geometry and/or Thomson scattering geometry.

CMEs identified in the LASCO data are tabulated in the CDAW (Co-ordinated Data Analysis Workshops) CME catalogue (Gopalswamy et al., 2009; https://cdaw.gsfc.nasa.gov/CME_list/). Again, the CDAW cataogue includes observational parameters such as time of first observation, central PA and PA width, but also derived information including speeds, acceleration and mass. The radially-restricted fields of view of a coronagraph (relative to the HI field of view) are not conducive to derivation of 3D information for observations from a single vantage point in the manner described above, so speed profiles are derived in the plane of the sky (whereas the HIGeoCAT speeds are radial speeds of the CME apex).

The CDAW catalogue does not list estimated CME onset times, but these can be derived by back-projecting to the solar limb using information in the plane of the sky. We estimated this for the CMEs studied in 2011 as part of the present analysis, as detailed in section 3. We note that from the LASCO vantage point, at L1, Earth-directed CMEs are seen as halos, as mentioned above, with associated ambiguities in determining speed, angular width and topology.

CMEs visually identified in the STEREO COR-2 data are listed in the Johns Hopkins University Applied Physics Laboratory (APL) catalogue at http://solar.jhuapl.edu/Data-Products/COR-CME-Catalog.php (Vourlidas et al., 2017). In addition to the spacecraft identification, an event ID and date and time of the first observation, each CME is classified by morphology with an established set of event-types, including flux-rope, loop, jet etc. Parameters such as central position angle, angular width and speed are also included, though, as with the CDAW catalogue, a large number of events are weak and, thus, the catalogue does not include entries for these parameters. Entries from the APL catalogue are compared to the HIGeoCAT and CDAW entries for each event studied.

The catalogues mentioned above enable comparisons to be made. They either provide estimated CME onset times or enable onset times to be estimated, assuming constant speeds. They also enable us to estimate CME arrival times at 1 AU, again also assuming constant speeds. Whilst we use such onset and 1 AU arrival times in this study, we note that there is much evidence for CME acceleration in both coronagraph and HI observations (e.g. Barnes et al., 2020; Ravishankar et al., 2020). Although most extreme acceleration occurs in the corona, there is evidence for acceleration and deceleration in the heliosphere, consistent with the influence of drag due to the background solar wind. This is particularly the case for fast or slow CMEs (see e.g. Harrison et al., 2012; Lugaz and Kintner, 2013; Vrsnak et al., 2013). The techniques used in this study do not take acceleration or deceleration into account because the work requires identification of likely associations rather than achieving the most accurate comparison of modelled and recorded ICME arrival time.

Any CME catalogue derived from the manual identification of CMEs is naturally subjective, and this applies equally to the CDAW, HIGeoCAT and APL catalogues. Events are identified by eye. Different observers may identify different events, and even having identified the same event, may characterize them differently. This is the case even when faced with identical imagery, especially at the weaker end of the brightness scale. In truth, no catalogue of this kind can be considered to be fully complete, and the comparisons being made in this work help us to understand this from an operational perspective. However, we note, as stated above, that HICAT includes 159 and 148 CMEs in 2011 for STEREO-A and STEREO-B, respectively, and, 116 and 122 were then analysed for inclusion in HIGeoCAT. For the same period, the CDAW catalogue includes 1947 LASCO CMEs and the APL catalogue lists 1599 COR-2 CMEs. The numbers are significantly different. One contributor to this is the fact that whilst the two coronagraphs cover full-revolution fields of view, covering all position angles, the HI-1 images are centred on the Sun-Earth line with an opening angle of about 70 degrees. Thus, the HI observations will not image back-sided CMEs (with respect to the Earth) or higher latitude CMEs. In addition, the two coronagraph catalogues do tend to include numerous small, weak outward propagating phenomena, whereas the HICAT catalogue is restricted to events with PA width in excess of $20^{o}$ (Harrison et al., 2018) in order to avoid the inclusion of the many blob-like structures moving through the inner heliosphere, which, most likely relate to the smaller, weaker events in the CDAW and APL catalogues. In this respect, the CDAW catalogue, for example, has a large number of events, most of which are labelled with words such as ‘extremely poor’, or ‘poor’. The classification of quality in the CDAW list is restricted to a comments column, with subjective and variable text, making it difficult to select, say, a full set of good or fair events to compare to the STEREO data.

There has always been an open debate about what classes as a CME when you consider the smaller events, but the ‘typical’ loop-like eruptions that one regards as classic CMEs tend to be clear in all three catalogues. The HIGeoCAT CMEs are those that are clear loop-like eruptions, classed as good or fair, i.e. those that we are confident that any observer would identify as a CME.

## 3 Analysis of the 2011 CMEs: Overview

Any study of coronagraph and heliospheric imagery of Earth-directed CMEs from STEREO over a prolonged period of time is complicated by the drifting configuration of the mission. For the purposes of this paper, we base our analysis on the HIGeoCAT catalogue from STEREO-A, in order to identify events of interest, for which we then consider observations from the other STEREO spacecraft (specifically HI-B and COR) and relevant SOHO data.

For each of the 116 HIGeoCAT HI-A CMEs from 2011, we attempt to identify an associated CDAW LASCO entry. We do this by making comparisons of CME estimated onset times and locations. Whilst CDAW provides the central PA of a CME in the corona, the source region could be anywhere on the disc (near-side or far-side) along the line of the projected track, between where it crosses the limb (as viewed from SOHO) and Sun-centre. Of course, this ignores any near-Sun deflection. For a CME of speed 350-400 km/s, that would mean that a CME onset time, assumed to be located at the limb, could be ~30 minutes later than the actual onset, even assuming a constant speed (i.e. no initial acceleration). Similarly, CME speeds given for halo/partial halo events, cited in the CDAW catalogue are speeds of a front, seen in the plane of the sky, which bares little relation to the CME apex.

The HIGeoCAT CME onset time is actually the time at which the CME would have been at zero degrees elongation (i.e. Sun centre), as this is what is output from the multi-variant fit. In this study, we use the output radial speed to correct the HIGeoCAT onset time such that it corresponds to the time that the CME crosses one solar radius. Correction of the onset time does not need to assume plane of the sky propagation. The principal errors will be down to uncertainties in the tracking of the CME apex, to which we are applying the geometrical fit, and erroneous assumptions in the fitting applied (i.e. constant speed and direction, and self-similarly expanding morphology). The latter two are assumptions also made in generating CDAW (the CME is assumed to be a point source, propagating in the plane of the sky).

Of course, the assumption of constant speed early in the CME trajectory is erroneous, but identifying acceleration, or, indeed, deceleration, especially in the occulted region of the field of view of a coronagraph, is problematic. However, the aim here is only to identify events observed in the heliosphere and in coronal imagery that are likely to be associated. So, whilst we acknowledge that there are uncertainties in deriving onset times, and we are not expecting LASCO and HI onset estimates for the same events to be within minutes of each other; we believe that LASCO and HI events can be associated objectively by defining appropriate time windows. For the purposes of this analysis, we define CMEs as common between the CDAW and HIGeoCAT catalogues if the CDAW CME onset time is within +/- 2 hours of the HI-A onset (corrected to 1 Rs) given by the SSEF30 model. We believe that this is sufficient to account for most uncertainties, and thus will capture most associated observations of the same CME. For the rest of this paper, all HI onset and Earth-arrival times will be those derived from the SSEF30 technique unless otherwise stated.

## 4 Selection of candidate Earth-directed CMEs

Of the 116 HIGeoCAT HI-A CMEs in 2011, 79% (92 CMEs) had an associated CDAW LASCO CME onset within +/- 2 hours. Some 30% (35) of the 116 CMEs had a CDAW halo or partial halo CME onset within +/- 2 hours. The latter rises to 37% (43) for a +/- 3 hour window. This suggests that many potentially Earth-directed CMEs are readily identified in both on and off Sun-Earth line images.

HIGeoCAT CME source locations are inferred from the SSEF30-derived CME longitude and latitude of the CME apex, assuming no deflection takes place between launch and the altitudes sampled by HI. On the HELCATS website, the longitudes and latitudes are given in HEEQ (Heliocentric Earth Equatorial) coordinates (Thompson, 2006). For a consideration of CMEs likely to be Earth-directed one could define an Earth-facing region centred on the centre of the solar disc as viewed from Earth. This is best defined using HEE (Heliocentric Earth Ecliptic) coordinates (Thompson, 2006). Thus, in HEE coordinates we define what we will call the Target Zone, given as +/- 20$^{o}$ east and west of the point where the ecliptic crosses solar central meridian, along the ecliptic plane, and +/- 20$^{o}$ north and south of the point on the solar disc immediately facing the Earth.

If we believe that a CME from an assumed source region within the Target Zone is likely to impact Earth, then 28 (24%) of the 116 CMEs would be Earth-directed. Selection of the Target Zone as defined is somewhat arbitrary. A larger CME originating from a source outside this region could still impact Earth, while a small CME from a source near the edge of this range could still miss Earth. In addition, we note the potential uncertainties in the estimation of CME source location using the SSEF30 method. However, we are simply looking for a set of likely Earth-impacting CMEs and believe that the defined +/- 20$^o$ region would provide us with such a set.

Of the 28 HIGeoCAT HI-A CMEs with inferred source regions within the Target Zone, only one had no associated CDAW CME onset (the nearest CDAW CME had a projected onset some 7.5 hours prior to the HIGeoCAT event onset). For 13 of the 28 HIGeoCAT CMEs, CDAW reported either a partial or full halo CME with onset within the +/- 2 hour window. For each of the remaining 14 cases there is a CDAW CME onset within +/- 2 hours of the HI-A corrected onset, although none were classified as partial or full halo CMEs. Thus, in summary, of 116 HIGeoCAT CMEs catalogued in 2011, some 28 (24%) were determined to be likely to be Earth-directed CMEs by virtue of the CME direction included in HIGeoCAT and 15 of these (54% of the 28) show apparent inconsistency between the CDAW and HIGeoCAT identifications in terms of their potential for Earth-impacts (one with no CME within +/- 2 hours and 14 with no partial halo/halo within the same time window). This takes nothing away from the fact that there is a close observational association between CMEs observed in the corona and heliosphere, for many events, but it does suggest that we should investigate these Earth-directed events rather more carefully to understand the apparent discrepancy.

Whilst the term CME is used in this study for events imaged in both the corona and heliosphere, CMEs identified in in-situ data are commonly referred to as Interplanetary CMEs (ICMEs). In 2011, there were 32 ICMEs detected at L1, as reported by Richardson and Cane using observations from the NASA ACE spacecraft (see Cane and Richardson, 2003; Richardson and Cane, 2010; and the catalogue of near-Earth interplanetary CMEs since January 1996 at www.srl.caltech.edu/ACE/ASC/DATA/level3/icmetable2.htm; henceforth referred to as the Caltech ACE catalogue). For the same year, 19 ICMEs were recorded, at L1, in the ICME catalogue for the NASA Wind spacecraft (Nieves-Chinchilla et al., 2018; see the NASA Wind catalogue at https://wind.nasa.gov/ICME_catalog/ICME_catalog_viewer.php; henceforth referred to as the NASA Wind catalogue). Given such low numbers of identified ICMEs per year, the lack of consistency between the HIGeoCAT and CDAW observations for 15 potentialy Earth-directed CMEs in 2011, is very significant.

We also note the discrepancy in the numbers of ICMEs identified in the ACE and Wind catalogues. This is illustrated by the figures given later, where for the 15 CMEs studied in detail, 6 have no associated ICMEs recorded (from ACE or Wind data), 4 have an associated ICME recorded in both Caltech ACE and NASA Wind catalogues, but for 3 events associated ICMEs were only recorded in the Caltech ACE catalogue and for 2 events associated ICMEs were only recorded in the NASA Wind catalogue. It is beyond the scope of this paper to address the difference between ICME identification using the ACE and Wind data, although the identification of in-situ signatures of CMEs is fraught with challenges. However, for the purposes of this study we attempt to correlate catalogued ICMEs recorded by either ACE or Wind to CMEs catalogued from coronagraph and HI imagery. We do this by using the SSEF30-derived radial speed in HIGeoCAT to estimate an arrival time at L1 assuming that the CME propagated with that constant speed all the way out to L1. It should be noted that we do not cater for the assumed CME geometry in the L1 arrival time, i.e. we do not apply the correction of Möstl and Davies (2013). The aim, again, of the paper is to identify likely associations rather than perform detailed investigation of CME arrival. Hence, we use an extended window of +/- 48 hours in the following analysis. Having noted this, we point out that the estimated arrival times derived below differ, on average, by only 112 minutes from those listed in the HELCATS ARRCAT catalogue (ARRval CATalogue; https://www.helcats-fp7.eu/catalogues/wp4_arrcat.html) that does correct for the CME curvature using the method of Möstl and Davies (2013). Thus, our analysis, below, should indeed be adequate for association of events.

## 5 The CME of 14 February 2011

Before examining the discrepant events, we show a canonical Earth-directed CME revealed in the STEREO and LASCO imagery and with a resultant signature observed in-situ near Earth. That said, even this event is still far removed from an ideal single Earth-directed CME. We choose the first halo CME of 2011, launched on 14 February, as reported in the CDAW list, for which there is a corresponding HIGeoCAT CME.

The CME in question first entered both STEREO-A and STEREO-B HI-1 fields of view at 22:49 UT on 14 February. SSEF30 analysis of the STEREO-A HI data reported in HIGeoCAT yields a radial speed of 415 km/s, an onset (corrected to 1 Rs) of 19:05 UT, and imply a source location (assuming no deflection during its passage through the corona) of $18^o$ longitude and $0^o$ latitude (HEEQ; equivalent to $18^o$ and $6^o$ HEE), which is within the specified Target Zone. HI-1 difference images of the CME at 17:29 UT on the following day are shown in Figure 1. Difference images (in which the previous image is subtracted from the current image) highlight enhancements and depletions in brightness relative to the previous image, shown as white and black regions, respectively, and, as such, largely removes the contribution of, for example, the slowly varying F- and K-corona. The high degree of symmetry between CME signatures in the HI-1A and HI-1B images suggests that the CME is close to Earth-directed, given the longitudes of the spacecraft ($-94^o$ for STEREO-B and $+87^o$ for STEREO-A). However, the HIGeoCAT SSEF30 entries derived from the HI-B data suggest a radial speed of 523 km/s, an onset (at 1 Rs) at 21:09 UT on 14 February from a longitude of $-54^o$ and a latitude of $21^o$. Such a direction is inconsistent with the results derived from HI-A data (and the symmetry of the images). This demonstrates clearly the inherent uncertainties of geometrical modelling techniques.

The CDAW catalogue includes two halo CMEs on 14-15 February. The first one is listed as a poor event, first observed in LASCO C2 at 18:24 UT on 14 February 2011, with an estimated speed of 326 km/s and a back-projected onset time of 17:20 UT. This onset time is reasonably consistent with the HI-A HIGeoCAT SSEF30-estimated onset of 19:05 UT, and less so, with the HI-B value of 21:09 UT, noting that these onset times do not account for acceleration or deceleration. The event can be identified through weak outflows within streamer structures around a wide range of PAs.

The second CDAW CME, on 15 February, was first observed at 02:24 UT in the LASCO images. This was a bright event with an estimated speed of 669 km/s. Although this CME was estimated to have launched much later than the onset time estimated from HI-A data, by the time that the two CDAW CMEs were in the field of view of the HI instruments the two could no longer be readily distinguished. Whilst Figure 1 shows images of the CME as viewed by the HI-A and HI-B instruments, the accompanying CDAW image shows the second of the LASCO CMEs.

Temmer et al. (2014) identified flare associations with the two CDAW-identified halo CMEs, namely, an M2.2 flare at S20 W04, with onset at 17:20 UT on 14 February, and an X2.2 flare at S20 W10, with onset at 01:44 UT on 15 February. These locations are consistent with the SSEF30 estimate of the HI-A CME source location, given above. Temmer et al (2014) suggest that there is an interaction between the two Earth-directed CMEs, and they also identify signatures in the HI data that they assign to both CDAW CMEs. However, HIGeoCAT only identifies one CME, which appears to be a merged structure resulting from the interaction of the two events imaged by LASCO. For more details of the CME-CME interaction, the reader is referred to Temmer et al. (2014).

Assuming no acceleration or deceleration beyond the track of the CME, the SSEF30 speed derived from the HI-A data of 415 km/s suggests an arrival in the vicinity of Earth at 23:01 UT on 18 February. The Caltech ACE catalogue reports the in-situ arrival of a CME at the ACE spacecraft at 19:00 UT on 18 February, with a speed 470 km/s; consistently, the NASA Wind catalogue lists a CME arrival at the Wind spacecraft at 19:50 UT on 18 February, with a speed of 462 km/s. Thus, the SSEF30-projected CME arrival time is within 3-4 hours of the ICME arrival times reported for ACE and Wind.

We note that both Mishra and Srivastava (2014) and Maricic et al (2014) have examined the same sequence of events and, indeed identify a third CME that they believe also interacted with the two CMEs discussed above before arriving at L1. They demonstrate that signatures of all three events can be seen in-situ at the Wind spacecraft.

Notwithstanding the inconsistency of the HI-B HIGeoCAT SSEF30-derived CME characteristics, and the complexity resulting from the presence of multiple halo CDAW events, this sequence appears to show Earth-directed halo CMEs, as imaged by the SOHO LASCO instrument at L1, that are correspondingly well imaged from an off the Sun-Earth line perspective, by the STEREO HI-A and HI-B instruments. It also demonstrates that the SSEF30 analysis of the HI-A data appears to predict an ICME arrival at Earth that is consistent with an actual ICME arrival recorded by ACE and Wind. That said, this discussion also demonstrates that, by necessity, catalogues can miss to pick up on the complexity that is revealed through in-depth studies of single events.

Having shown evidence for one event sequence that is a reasonable fit to the established view, albeit with some complexities, we now examine the 15 CMEs for which a potentially Earth-directed CME is identified in the HIGeoCAT catalogue but with no consistent CDAW CME.

## 6 Analysis and discussion of the discrepant events

Table 1 includes salient parameters of the 15 discrepant CMEs identified by the method discussed in section 4, based on the HI-A SSEF30 entries in the HIGeoCAT catalogue. For each of the HI-A-observed CMEs, we identify any corresponding HI-B CME, using the HELCATS HIJoinCAT catalogue, to be the same event observed from the other STEREO spacecraft. Thus, in Table 1, we include the HI-B entries next to their HI-A counterparts, where possible. For some events, no corresponding HI-B CME is recorded in the HIGeoCAT catalogue.

Columns 2 to 9 of Table 1 list a number of HI CME parameters from HIGeoCAT. The HELCATS event ID (column 2) is a unique identifier for each HICAT/HIGeoCAT CME, consisting of (1) the STEREO spacecraft from which the CME was detected (i.e. STEREO A or B), (2) the date of entry of the CME into the HI-1A or B field of view and (3) an additional index to legislate for the fact that multiple CMEs can potentially enter the same HI-1 field of view on the same date. Column 3 includes the date and time that the CME is first observed in the HI-1A or B field of view. As noted above, the SSEF30 technique provides an estimate of onset time at Sun-centre (column 6) from which we derive a corrected onset time at 1 Rs (column 7); the CME source location is inferred (assuming no coronal deflection) from the CME apex latitude/longitude (column 4 and 5; HEEQ coordinates). We also reproduce the equivalent HEE coordinates (also included in column 4 and 5).

The table also includes the SSEF30 estimate of the CME radial speed (column 8), and (derived assuming tha CME propagates at that speed out to 1 AU) the estimated arrival time calculated at 1 AU (column 9). The geometrical analysis techniques used to generate HIGeoCAT is described by Barnes et al. (2019). However, in deriving the estimate of CME arrival time at 1 AU, it is useful to discuss briefly the uncertainty on the estimated speed from HIGeoCAT. HIGeoCAT lists an uncertainty in the speed for each event, which is based on the errors in fitting the profile of the CME front on a time-elogation map. For the events given in Table 1, that error is of order a few percent. However, of greater significance will be accounting for errors in the underlying assumptions of the technique (constant speed and direction, CME morphology), or errors in tracking the CME. This could vary significantly between events. However, because our aim is to identify associations rather than derive the most accurate arrival times, we use a generous time-window, as defined below.

Table 1 also lists key CDAW information, including the time of first CME observation by LASCO C2 (column 10), the back-projected CME onset (column 11) and CME apex speed (column 12), for all CMEs with back-projected onsets within +/- 2 hours of the associated HIGeoCAT HI-A CME onset  (corrected to 1 Rs).

Finally, Table 1 includes the arrival times and speeds of ICMEs  as recorded in-situ at the ACE (columns 13-14) and Wind (columns 15-16) spacecraft, extracted from the Caltech ACE and NASA Wind catalogues, respectively. The tabulated information for the two catalogues is slightly different and that is reflected here by the inclusion of shock arrival time, plus ICME arrival and end times from the ACE catalogue and the ICME arrival time, followed by ‘magnetic obstacle’ onset and end time from the Wind catalogue. Note that for a potentially associated ACE and Wind ICME to be included in Table 1, the projected CME 1 AU arrival derived from the SSEF30 results from HI-A data must be within 48 hours of the shock time (ACE), ICME arrival time (Wind) or ICME or magnetic cloud onset and end times (ACE and Wind).

From the 15 potentially Earth-directed CMEs (i.e. HI-A observed CMEs with source regions within the defined Target Zone) from 2011 without consistent listed CDAW entries, as listed in Table 1 we select two for which corresponding ICME arrivals are recorded at Earth, and describe them in some detail. These are the CMEs of 11 July and 25 May (events 10 and 5 from Table 1). We also describe the CME of 24 January (event 1), which is an example where a potentially Earth-directed CME, as observed using HI-A, was identified in the HIGeoCAT catalogue for which, there was no consistent CDAW CME and no corresponding near-Earth ICME. The remaining events for which there was no corresponding Earth arrival are briefly summarised, afterwards.

**6.1 The CME of 11 July 2011 (CME number 10 in Table 1)**

Figure 2 shows the CME first identified in the HI-1A and HI-1B images on 11 July 2011. A sequence of selected near-simultaneous HI-1A and HI-1B difference images are displayed in Figure 2.

Whilst it appears that HI-1A and HI-1B are imaging the same CME from different vantage points (in this case 190$^{o}$ apart), it is noted that, in Table 1, the estimated HIGeoCAT SSEF30 onset times are over 4 hours apart. Both HI-1A and HI-1B difference images show an ill-defined loop-like structure, where the amorphous nature of the bright loop-front could well lead to difficulties in accurately tracking the same feature between images to obtain the most comparable SSEF30 results; this is a possible explanation for the difference in onset times derived from the two spacecraft. Both HI-A and HI-B images show that the CME is directed slightly south of the ecliptic plane. There also appears to be evidence for a previous, narrow (in PA extent), CME propagating well north of this CME. SSEF30 analysis, assuming no CME deflection sunward of the HI field of view, places the CME source regions derived from HI-A data and HI-B data, at sites just 14$^{o}$ apart in HEEQ longitude and 1$^{o}$ apart in HEEQ latitude; the HI-A source region is well within the Target Zone and the HI-B source region is just outside.

There are two CDAW CMEs (see Table 1) with back-projected onset times within 2 hours of the estimated CME onset from HI-A data. These CMEs were both noted as being 'very poor' events, with neither classified as a halo or partial halo event. They both consist of outflows associated with streamers projected as lying off the western solar limb and, as such, do not appear to be consistent with the loop-like CME originating on the Earth-facing hemisphere of the Sun, imaged in the HI data. The CDAW events also displayed speeds some 200 km/s lower than those from HIGeoCAT. It is possible that there is an Earth-directed CME that is too weak to be identified as such, for which the flows in the streamers are signatures of the passage of the flanks of the CME, with speeds that are associated with the expansion of the flanks rather than the speed of the CME apex. Whether this is the case, or not, we conclude that the CDAW catalogue does not identify the HIGeoCAT CME that is potentially Earth-directed.

The COR-2 APL catalogue lists a CME first seen in COR-2A at 11:24 UT on 11 July and in COR-2B at 11:54 UT, propagating just south of the equator on the Earth-facing side of the Sun. The timing and structure, including propagation direction, of the CME (Figure 2), and its speed (plane of sky speeds given in the APL catalogue), at 445 km/s and 449 km/s, in COR-2A and COR-2B, respectively, appear to confirm that it is the same near-Earth directed CME identified in the HI data.

Thus, from the off Sun-Earth line platform we have an identification of a potentially Earth-directed CME, that is not identified as such in the CDAW catalogue.

The Caltech ACE catalogue records the arrival of an ICME at L1, on July 14 (Table 1). There is no corresponding event in the NASA Wind catalogue. Table 1 shows good consistency between the HI CME projected arrival time and the actual ICME arrival time at ACE in that the timing of the ICME shock and the ICME onset and end encompass the projected CME arrival times from the HI -A and HI-B data, and the in-situ speed, at 410 km/s, is not inconsistent with the SSEF30 CME speeds of 494 km/s and 361 km/s derived from the HI-A and HI-B data, respectively.

There were no other CMEs identified in the HIGeoCAT catalogue that could have resulted in the arrival of a CME in the vicinity of the Earth in this timeframe. Thus, we conclude that the off Sun-Earth line observations from HI and COR-2 appear to reveal an Earth-impacting CME that was not identified as such in the CDAW catalogue and not identified by re-examination of the LASCO data.

### 6.2 The CME of 25 May 2011 (CME number 5 in Table 1)

Figure 3 shows a selection of HI-1A and HI-1B difference images from the CME that was first observed in the HI data on 25 May 2011, in the same format as Figure 2. The HI CME can be seen clearly, as a loop-like CME with a concave front. As cited in Table 1, this CME was first seen in the HI-1A and HI-1B fields of view only 40 minutes apart and the estimated source locations, from the HI-A and HI-B observations, are within $17^{o}$ in HEEQ latitude and $3^{o}$ in HEEQ longitude of each other, both within the +/- $20^{o}$ Target Zone. The radial speeds derived from HI-A and HI-B are also consistent with one another (being 47 km/s apart). The estimated HI-A and HI-B onset times of the CME (at 1 Rs) are about 2.5 hours apart, which is larger than one might expect (the difference in the speed would only account for a 15 minute discrepancy), but this is a consequence of the fact that the analysis proceeds by means of multi-variant fitting. It is clear that the two instruments are imaging the same CME.

Figure 3 also shows a following CME, a bright loop-like event just entering the fields of view of HI-1A and HI-1B at the start of the sequence, and overtaking (in terms of elongation) the northern portion of the first CME. That second CME was first observed at 14:49 UT (15:29 UT) in the HI-1A (HI-1B) field of view, had an estimated speed of 1026 km/s (1181 km/s), and an estimated onset at 11:45 UT (11:02 UT), from a source at $-21^{o}$ ($41^{o}$) HEEQ longitude and $6^{o}$ ($11^{o}$). HEEQ latitude. Since its estimated source location is outside the specified Target Zone for both HI-A and HI-B SSEF30 analysis, this second CME is excluded from Table 1.

The only CDAW CME with a back-projected onset within 2 hours of the HI-A CME 1 Rs onset time had an onset time some 53 minutes after the HI-A onset. The CDAW and HIGeoCAT speeds are inconsistent in that the latter are around twice as fast as the listed CDAW CME speed, though this could well be a projection effect due to the former being in the plane of the sky. The CDAW CME (as shown in the LASCO C2 image in Figure 4) is identified as a $38^{o}$ wide outflow of material along a streamer centred on the solar north-west limb at PA $328^{o}$ (as viewed from SOHO). The CDAW CME was not classified as either a full halo or a partial halo CME, and was recorded as a ‘very poor event’.

The COR-2 APL Catalogue lists a CME first detected in COR-2A images at 04:39 UT and in COR-2B images at 04:54 UT. The bottom panel of Figure 3 shows the COR-2 CME as a loop-like event off the eastern limb (Earth-facing side) of the Sun, with location, timing and morphology consistent with the HI CME.

As with the previous event, we have a CME imaged in the heliosphere, by the STEREO HI instruments, that is identified as being potentially Earth-directed, and that is consistent with the coronal images from COR-2, but with no corresponding Earth-directed CME recorded in the CDAW catalogue.

The projected 1 AU arrival times of the CME based on the estimated speed from SSEF30 analysis of HI-A and HI-B data are consistent with entries in both the CALTECH ACE and NASA Wind ICME catalogues. The HI CME speeds are also consistent with the catalogued speeds recorded at 1 AU. This provides some confirmation that the off-Sun-Earth line observations clearly showed an Earth-impacting CME that was not identified from L1 coronagraph data.

To support or refute this conclusion in this case, we need to consider a few additional points.

As discussed above, we identified a second apparently faster CME. Could the ICME have been due to that second CME, rather than the first CME that is listed in Table 1? The second CME had an estimated onset some 8 to 10 hours later than the first CME, but it appeared to overtake the first CME. While the second CME could have passed either in front or behind the first CME, the two could have interacted even if only partly intersecting. The HIGeoCAT entries for the second CME imply that its apex is at around $20^{o}$ higher latitude than that of the first CME, but the HI-A and HI-B HIGeoCAT apex longitude estimates differ significantly from each other, at $-21^{o}$ and $41^{o}$, respectively. This discrepancy could be at least partly due to inaccuracy in tracking the second CME resulting from the superposition of the two CMEs. In trying to identify a likely source region for the second CME, we note that the CDAW catalogue lists a LASCO CME with an onset time of 12:40 UT, consistent with that of the second CME, and a plane of sky speed of 561 km/s. This CDAW CME is listed as a poor event, and not as a halo or partial halo CME but, rather, is centred in the solar north-west. There is no significant flare activity within 5 hours of the estimated onset time of this second CME. There is no obvious solar source for this CME but, between 08:00 and 12:00 UT the SDO images show a filament eruption at high latitude in the solar north-east quadrant, as viewed from Earth. Whether they are connected or not, the filament and CDAW CME demonstrate that there is evidence for eruptive events in the corona at higher latitudes than the first CME in the time-frame of the second CME, apparently across a wide longitudinal range. This could contribute to the discrepancy between the HI-A and HI-B source locations.

Thus, whilst we cannot be sure about the longitude of the second CME, there are indications that it is not likely to have been Earth-directed and that its direction of propagation was at higher latitude. Even so, we cannot rule out any interaction between the two CMEs. In any case, though, there is no suggestion from the CDAW entries that either is an Earth-directed CME.

If the second CME had continued at a constant speed, it would have arrived at 1 AU at 04:22 UT on 27 May and 22:19 UT on 26 May, based on HI-A and HI-B speed estimates, respectively. Neither ACE nor Wind in-situ measurements report an ICME arrival for some weeks prior to the catalogued ICME arrival on 28 May, and that ICME arrival is, as discussed above, consistent with the estimated arrival of the first STEREO-imaged CME (which is listed in Table 1).

Our conclusion is that the HI and COR-2 observations show evidence for an Earth-directed CME that did appear to arrive at Earth, as evidenced by in-situ measurements, whilst the CDAW catalogue reports no evidence of a corresponding Earth-directed (halo/partial halo) CME.

### 6.3 Other suspected Earth-impacting CMEs with no associated CDAW events

There are 15 CMEs listed in Table 1. Nine of them, including the two discussed above, correspond to cases where HIGeoCAT provides evidence for an Earth-impacting CME with no associated CDAW full or partial halo entry, but that is consistent with the detection of an ICME arrival at 1 AU, at ACE or Wind. We briefly summarise the remaining cases here and images for a selection of these events are shown in Figure 5.

**CME 8 (23 June 2011):** The 23 June 2011 CME consists of a bright multiple-loop event that displays symmetry between the HI-1A and HI-1B images. Using the SSEF30 technique the estimated source locations are within the Target Zone for both the HI-1A and HI-1B data. The SSEF30 estimated speeds differ by 115 km/s resulting in estimated onset times about 4 hours apart, but the images clearly demonstrate that the instruments are imaging the same CME. One LASCO CME is identified within +/- 2 hours of the estimated onset from the HI-1A data. Centred on position angle 341$^{o}$ in the solar north-west, it is not listed as a partial halo or halo event but is a narrow event associated with the bright streamer in the north-west. Whereas there is no reported ICME in the Caltech ACE catalogue, there is an ICME listed for the NASA Wind catalogue that first arrives at L1 almost half a day after the SSEF30-estimated arrival of the HI-1A CME, assuming constant speed. Thus, this event appears to be a CME that is likely to be Earth directed, but with no corresponding entry in the CDAW catalogue, and a consistent ICME arrival recorded in the vicinity of Earth.

**CME 9 (03 July 2011):** This is a multi-loop-like CME in the HI images (Figure 5, top row), although the images are complicated by the remnants of a prior CME to the north. Indeed, whilst the HI-B images show CME 9, as well as features relating to the prior CME, CME 9 is not listed in HIGeoCAT. This is most likely due to the fact that the CME is fainter in HI-B images than in the HI-A images. The only CDAW-listed CME potentially associated time-wise with this event was inconsistent in terms of speed and location (even taking into account the different observing geometries). Not listed as a halo or partial halo, it was identified as being centred in the solar south east, and was classed as a very poor event. The estimated CME 1 AU arrival time, based on the SSEF30 analysis of HI-A imagery, is approximately 21 hours prior to the arrival of an ICME at Earth, as listed in the Caltech ACE catalogue. This arrival time discrepancy can be explained by the difference between the SSEF30 speed (713 km/s applied to HI-A) and the much-lower in-situ speed (360 km/s), i.e. due to deceleration beyond the limit of the HI track, as might be expected for such a fast CME. An ICME recorded in the NASA Wind catalogue, that arrived at the spacecraft on 3 July is considered to be too early to be a signature of this event, arriving as it did only 23 hours after the onset time of the CME estimated from HI-A data. In any case, there was no CDAW halo or partial halo candidate for either of the two ICME arrivals referred to.

**CME 11 (05 October 2011):** In the HI images, this CME is also complicated by the remnants of an earlier CME, to the north. Whilst CME 11 is included in HICAT (and thus HIGeoCAT) for HI-A there is, again, no corresponding entry in HICAT/HIGeoCAT for HI-B. That said, HI-1B images do appear to show, albeit less clearly, the same CME. The CME is also evident in the COR-2 images. The two CDAW-listed CMEs (Table 1) with back-projected onset times within 2 hours of the HI-A CME estimated onset time are not consistent with the HIGeoCAT CME in terms of morphology and location (even taking into consideration the different observational geometry). They were not classified as halo or partial halo CMEs, were both described as poor events corresponding to narrow streamer outflows. The estimated 1 AU arrival time of the HIGeoCAT CME, based on SSEF30 analysis of the HI-A imagery, is consistent with an entry in the Caltech ACE ICME catalogue for 7 October.

**CME 12 (26 October 2011):** The HI images for this CME reveal a complex loop-like event (Figure 5, middle row). Whilst HI-A SSEF30 analysis suggests that this is a potentially Earth-directed CME, analogous HI-B analysis suggests a CME source on the East limb with respect to Earth (see Table 1). The CME is much clearer in the HI-1A images, than in simultaneous HI-1B images, and the COR-2A observations show a CME that is consistent with that seen in HI-1A. We consider this to be a potentially Earth-directed event, whilst noting the inconsistency in the HI-1B SSEF30 analysis.

There are two LASCO CMEs included in the CDAW list with back-projected onset times within 2 hours of the estimated HI-1A CME onset time. In common with all of the events discussed in Section 6, these were not classified as halo or partial halo events, and were both noted as being poor. Both CDAW CMEs were characterised, again, as narrow outflows along streamers off the solar south-east limb. Neither event appears to be consistent with the topology and estimated direction (assuming radial propagation from the estimated source location) of the HI CME.

The projected CME arrival time at 1 AU, and the HI-A SSEF30 CME radial speed used to derive that arrival time, are remarkably consistent with the values for this ICME in the Caltech ACE catalogue entry for 30 October. Thus, it appears that the HI-A and COR-2 data showed an Earth-directed CME, which indeed impacted the Earth, but there is no corresponding halo/partial halo CDAW CME. We noted that there is some inconsistency with the HI-B event, which could be due to lack of clarity of the images, leading to inaccuracy in the tracking, or, indeed, to the possibility that another CME is being witnessed in the HI-B images.

**CME 13 (29 October 2011):** HI and COR-2 images for this event exhibit a very similar morphology to CME 12. The two CDAW-listed LASCO CMEs that have back-projected onset times within two hours of the estimated HI-A CME onset time were both identified as poor events, and neither classed as a halo or partial halo CME. Again, the CDAW CMEs were narrow events with topology and apparent directions of motion that appear to be inconsistent with that of the HI CME. They were not identified as signatures of an Earth-directed CME.

Both HI-A and HI-B SSEF30 analyses provide estimated 1 AU arrival times on 01 November, which is consistent with an ICME entry for 01 November in the Caltech ACE catalogue and also one in the NASA Wind catalogue. An earlier ICME listed in the Caltech ACE catalogue on 30 October could conceivably be related, but it is much better correlated with CME 12. Also, in the NASA Wind catalogue there is an ICME that arrived later on 2 November, which again could conceivably be related to CME 13, but the timing of the 01 November arrival is most consistent. Either way, it is likely that HI and COR-2 detected an Earth-directed CME that indeed arrived in the vicinity of the Earth, but was not identified from the CDAW-listings of LASCO CMEs.

**CME 14 (01 November 2011):** This CME is seen as a bright loop emerging from equatorial latitudes in both HI-1A and HI-1B images, though it was classed as a poor event in HI-1B and, thus, not included in HIGeoCAT. A preceding CME is evidenced as a large loop like structure propagating ahead of CME 14. The COR-2 images appear to show the same CME. However, it appears to be directed slightly more northward than the HI images would indicate, though that could be evidence for a subsequent equatorward deflection (see e.g. Kahler et al., 2012; Harrison et al., 2012).

There is one CDAW-listed LASCO CME with a back-projected onset within 2 hours of the HI-A CME onset time on 01 November. It was not classed as a halo or partial halo event, and was described as poor. The CDAW CME was listed as a very faint structure flowing through or adjacent to a streamer off the solar north-east limb as viewed from L1. The topology, apparent direction of motion and timing of this CME did not lead to an identification of a CME corresponding to that imaged in the HI and COR-2 data.

The projected 1 AU arrival time of the CME, early on 05 November, derived from results of application of the SSEF30 technique to HI-A images and assuming constant speed propagation, was only six hours after the recorded arrival of an ICME in the NASA Wind catalogue. No associated event was recorded in the Caltech ACE catalogue.

The CDAW catalogue does include a partial halo event, first observed at 20:00 UT on 02 November, with a plane of sky speed of 384 km/s. With a back-projected onset time well over 24 hours later than the estimated onset derived using the HI-A data, we do not associate this partial halo with CME 14. However, we need to be sure that it is not related to the ICME arrival mentioned in the last paragraph. Based on the CDAW plane of sky speed, that CME would arrive at 1 AU at 08:50 UT on 07 November, over 2.5 days after the arrival of the ICME at L1, and over 1.5 days after the end of the event at L1. Thus, we conclude that the CME observed from STEREO is the most consistent CME to be associated with the 1 AU arrival on 4 November, and that this Earth-directed CME was not identified in the CDAW catalogue.

**CME 15 (29 November 2011):** The HI-1A images for this event reveal a complex, bright loop with the remnants of a preceding CME further out (Figure 5, bottom row). The single-spacecraft SSEF30 analysis of the HI-A bright loop suggests a potentially Earth-directed CME. The HI-B images appear to show the same CME, extending to roughly the same elongation at the same time, which is suggestive of an Earth-directed event, despite the fact that the estimated onset times differ by four hours (compared to the HI-1 entry times, which only differ by 1 hour). However, the structural complexity of the CME in question, plus the fact that it follows closely after another CME, could have hindered accurate tracking of the CME. Whilst analysis of the HI-A data suggests a CME source well within the target zone, the source for the HI-B data appears to lie just outside the Target Zone. The COR-2 images show the same CME earlier in its development. We believe that there is sufficient evidence to suggest that STEREO is imaging an Earth-directed CME.

The CDAW catalogue includes no LASCO CMEs at all with projected onset times within 2 hours of the HI-A-estimated CME onset time. Thus, the HI and COR-2 data reveal a potentially Earth-directed CME that was not identified in the CDAW catalogue.

The Caltech ACE catalogue includes an ICME arrival that corresponds to the SSEF30-estimated arrival time of the CME on 02 December; the ICME shock arrival is just over 6 hours prior to the estimated CME arrival.

We note that CDAW lists a LASCO partial halo CME, first observed on 29 November at 23:12 UT. Its back-projected onset time was considered to be outside the window of association for CME 15, but given its speed of 768 km/s, its arrival at 1 AU (given a constant speed) would be on 02 December at 04:40 UT. This is over 12 hours prior to the ICME arrival, but could make it a candidate for the ICME instead of CME 15. However, despite its classification as a partial halo, the LASCO images show a clear loop-like event confined to the solar south-western quadrant and that appears to match a HIGeoCAT CME with an estimated source region 60$^{o}$ west of central meridian. Considering this, we suggest that it is most likely that the CME of 29 November is an example where a potentially Earth-directed CME was detected in the HI and COR data, is not included in the CDAW catalogue, and later L1 in-situ observations suggest a consistent ICME arrival at Earth.

## 7 Discussion: Earth-impacting events

In all, 9 CMEs in 2011 have been described where off Sun-Earth line observations from the HI instruments aboard STEREO have imaged potentially Earth-directed CMEs, which appear to be related to subsequent ICME arrivals at Earth, but that are not identified as Earth-directed (halo/Partial halo) CMEs in the CDAW catalogue.

We do not claim that these are the only such events in 2011, because we have restricted our analysis to CMEs with apparent sources derived from SSEF30 analysis of HI-A data within the specified Target Zone. Events emerging from outside that zone could potentially also impact Earth, if sufficiently wide, for example. However, the SSEF30 analysis assumes a constant direction and speed fit to a circular cross-sectional geometry that will most likely have variable success, due to the level of compliance of each CME with these assumptions and the accuracy with which the CME fronts can be tracked in the time-elongation maps from which the time-elongation profile is extracted by hand. Difficulties in identifying and consistently tacking a particular CME track can be encountered due to superposed CME activity. However, the SSEF30 technique is designed to enable a simple method for event analysis. Nevertheless, the robustness of the SSEF30 method has been demonstrated previously (e.g. Mostl et al., 2022) and, despite the anticipated uncertainties, we believe that the general conclusions of the analyses discussed above are valid. Thus, it is believed that we have identified Earth-directed CMEs in the off Sun-Earth observations, that are not identified from L1.

## 8 Non Earth-impacting CMEs with no associated CDAW events

The starting point for Table 1 was to identify CMEs from 2011 in HIGeoCAT that had a reasonable chance to be Earth-directed, defined by the SSEF30-derived HI-A CME source being within what we called the Target Zone. Of the 28 HIGeoCAT CMEs that satisfied this criterion, we further identified those for which there was no associated CDAW halo or partial halo CME. Thus, we identified 15 so-called discrepant CMEs that were identified as being likely to be Earth-directed, from the HIGeoCAT HI-A catalogue, but had no report consistent with a likely Earth-directed CME from the CDAW catalogue.

As discussed above, of the 15 discrepant cases, 9 appear to relate to ICME arrivals at Earth as recorded in the Caltech ACE and NASA Wind ICME catalogues. Thus, we have 6 cases where potential Earth-directed CMEs were identified in the off-Sun Earth line data only, but which did not appear to result in an ICME arrival at Earth as noted in the ICME catalogues.

From a forecasting point of view, we need to be aware of all potentially Earth-directed events, i.e. all 15 are of interest. Below, we describe one sequence of events, that of 24

January 2011 (Table 1, CME 1), to illustrate one of the events that did not appear to arrive in the vicinity of the Earth. We then comment very briefly on the remaining 5 events.

### 8.1 The CME of 24 January 2011 (CME number 1 in Table 1)

Figure 6 shows the CME of 24 January 2011 (CME 1), as imaged by HI and COR-2. In the HI images, the CME is a large event on the Earth-facing side of the Sun, centred near to the ecliptic plane. It has a complex loop-like structure with a central position angle the ecliptic plane, with some extended structure to the north. Both HI-1A and HI-1B appear to image the same CME, and their slight asymmetry is indicative of a CME that could be directed somewhat westward of central meridian. The SSEF30-estimated 1 Rs onsets are 109 minutes apart, whilst they show similar radial speeds (380 km/s and 388 km/s; Table 1). The SSEF30-derived propagation directions are almost identical, apart by just $3^o$ in HEEQ latitude ($12^o$ and $15^o$) and $5^o$ in HEEQ longitude ($-6^o$ and $-1^o$), translating to an HEE source location for HI-A (HI-B) at $+12^o$ ($+14^o$) latitude and $0^o$ ($+5^o$) longitude. This suggests that the source is well within the Target Zone, just into the northwest quadrant of the solar disc, and potentially Earth-directed.

The APL COR-2 catalogue lists a CME first observed on 24 January at 02:39 UT (COR-2A) and 03:54 UT (COR-2B) with plane of sky speeds of 320 and 318 km/s, respectively, indicating consistency with the HI CME. The COR-2A image shown in Figure 6 indicates a similar CME structure to that seen in the HI images.

There was just one CDAW CME with a back-projected launch within 2 hours of the derived HI-A CME onset time. The next nearest CDAW CMEs estimated lauch times were over 5 hours prior to and over 11 hours after the estimated HI-A CME onset time. Figure 6 also shows a LASCO C3 image of this CME, which manifests as a loop-like event propagating to the south-west of the Sun, centred at position angle $260^o$ and of width $79^o$. Clearly this was not identified in the CDAW catalogue as a halo or partial halo event and would, thus, not be regarded as potentially Earth-directed.

Figure 6 also shows an image from the Atmospheric Imaging Assembly (AIA; Lemen et al., 2012) instrument aboard the Earth-orbiting NASA Solar Dynamics Observatory (SDO). The image is over exposed to show an erupting prominence on the south-western limb at 01:00 UT. Activation of this southern polar crown prominence started early on 24 January, with the eruption clearly underway by 01:00 UT. The eastern leg of the prominence was well onto the solar disc with the structure extending beyond the limb at about position angle $230^o$. Indeed, after the eruption, the bright prominence structure ascended in a non-radial motion, centred on $230^o$ by about 02:21 UT. The CDAW-listed LASCO CME showed core-structure consistent with prominence material, and continued non-radial (equatorward) motion could result in the prominence being consistent with the LASCO CME. There is no other candidate prominence eruption in the south western quadrant of the Sun.

Whilst the LASCO CME, and the SDO prominence do not appear to be obviously related to an Earth-directed CME, there are features in the LASCO, COR-2 and HI images that could suggest that all three are showing the same event, or parts of the same event. If this is the case, it seems that at least a large portion of the eruption emerges from the south-western quadrant of the Sun. However, the HI-A and HI-B data suggest that at least a significant part of the eruption is directed towards Earth, and that was not noted in the CDAW catalogue. In short, on the basis of the HI and COR-2 data, from two spacecraft, an Earth-directed CME would have been forecast.

Whilst Table 1 shows an estimated arrival time at Earth, the Caltech ACE and NASA Wind catalogues do not record any CME arrivals at L1 between 24 January and 4 February 2011. Thus there is no recorded CME arrival at L1 that matches the CME events in question. Of course, the CME could have passed near Earth, but we do not have any evidence for this.

### 8.2 Other non-impacting events

The remaining CMEs listed in Table 1, not mentioned specifically above, were also cases where the HIGeoCAT catalogue indicates a potentially Earth-directed CME with no clear corresponding CDAW halo/partial halo CME, but there was also no associated ICME arrival in the vicinity of the Earth recorded in either the Caltech ACE and NASA Wind catalogues. This includes the events of 14 and 20 March (events 2 and 3), 3 May (event 4), 4 and 6 June (events 6, 7), 2011.

## 9 Discussion

In this study, our aim was to assess our ability to detect Earth-directed CMEs through the comparison of existing catalogues for both L1 and off Sun-Earth line observations. Particular focus was given to the comparison of the SOHO LASCO CDAW catalogue and the STEREO HI HIGeoCAT catalogue. For the latter, we had the choice of using three different geometrical models, as described in section 2, namely, the Harmonic Mean (HMF), Fixed-Phi (FPF) and Self Similar Expansion (SSEF; specifically the SSEF30 technique as described above) fitting techniques. For our comparison, we could have chosen to use any of these techniques and, indeed, it is often the case that they produce fairly similar results, particularly for certain geometries (including the geometry of an Earth-directed CME viewed from near L5). That said, the FPF technique tends to produce a slower speed and earlier estimated onset time than the SSEF30 technque, and the HMF technique tends to produce the fastest speeds and later estimated onset times. However, the differences are not, for the orbital geometry of this study in particular, significant enough to negate the windows of associaton that are used. If the study had been aimed at comparing CME substrucures with their in-situ counterparts, for example, then we would indeed require a more detailed analysis and arguably more sophisticated models.

One issue that must be addressed is why, from an observational point of view, we might be unable to identify Earth-directed CMEs from an L1 vantage point, that can be seen from off the Sun-Earth line. An issue that is related to the detection of halo CMEs concerns the so-called Thomson surface. Due to the Thomson scattering process, scattered photospheric emission is maximised at 90$^{o}$ to the line on the incident radiation, thus defining a sphere with diameter linking both the observer and the Sun (see e.g. Vourlidas and Howard, 2006; Howard and DeForest, 2012). This has been interpreted to mean that, for coronagraph observations, a CME directed towards Earth, which is propagating away from the Thomson surface, would be less bright than a similar CME in the plane of the sky (near to the Thomson surface). However, the effect has been shown to be less significant than earlier authors believed, leading to the use of the term Thomson Plateau (see Howard and DeForest, 2012, and references therein). From a geometrical point of view, there are other issues that reduce the visibility of CMEs when observed head on (as a halo or partial halo). The early propagation of a CME, especially a narrow, Earth-directed CME, is obscured by the occulting disc of the coronagraph, meaning that the CME is not within the field of observation until it is inherently much weaker. In addition, the intensity of a CME, which is integrated along the line of sight, is most likely, generally weaker for an event imaged head on, simply due to the geometry and mass distribution of the CME.

An addition issue that appears to hinder the recognition of a halo or partial halo CME, as indicated in a number of the events discussed above, is due to the superposition of pre-event or adjacent CME material, or multiple CME events, where the single L1 direction of observation cannot readily distinguish the Earth-directed halo CME effectively.

This study appears to reveal inconsistencies that have not been widely highlighted before. One relevant study is that of Möstl et al. (2014), who performed a study of 22 spacecraft-directed CMEs from the period 2008 to 2012. They used geometrical modelling of HI-imaged CMEs to predict speeds and arrival times at 1 AU, either at one of the STEREO spacecraft or at the Wind spacecraft, for comparison to in-situ plasma observations made at those spacecraft. Of the 22 CMEs, 19 were detected at the Wind spacecraft and were, thus, Earth-directed. The authors do not claim to have produced a list of all Earth-directed CMEs during their study period, but it is relevant to compare their study with the current work because their study period covers the whole of 2011. Möstl et al. (2014) selected events on the basis of clear in-situ signatures at 1 AU (at the STEREO or Wind spacecraft) and having clear HI time-elongation maps encompassing the associated CMEs. In this work, we simply identify all potentially Earth-directed CMEs from HIGeoCAT HI-A entries, irrespective of the presence of in-situ ICME arrivals. Thus, the current study compliments that of Möstl et al (2014) rather than duplicates it, and we should not expect the event lists included in the two papers to be identical. However, it is worth noting that, of the 15 discrepant events discussed, none was included in the Möstl et al. (2014) study.

For a number of these events, as described above, there were observational complexities, such as remnants of preceding CMEs, and these may have contributed to a lack of clarity in the time-elongation signatures that would most likely have resulted in such events not being considered in the Möstl et al. (2014) study. In addition, whilst the current study makes use of the ACE Caltech and NASA Wind ICME catalogues, the Möstl et al. (2014) study only employed Wind ICME data.

Similarly, Braga et al. (2020) produced a study of time of arrival in the vicinity of the Earth of CMEs imaged in HI data. Their initial CME list comprises 38 events studied by Sachdeva et al (2017) and it includes 13 HI-detected CMES from 2011. We note that only two of those 13 CMEs are listed in Table 1, above. The first of these two was the 24 January 2011 CME (CME 1). The Sachdeva et al (2017) study was concerned with the relative contributions of Lorentz forces and aerodynamic drag to CME propagation; they did not make a prediction of time of arrival at Earth. In Table 1 of Braga et al. (2020), they indicate that there is an ICME associated with the 24 January CME. However, the ICME 1 AU arrival time is given as 24 January 06:43 UT, which is prior to the first observation of the CME in the HI field of view, and Braga et al. (2020) did note that the timing (CME observation to ICME arrival) was too short. Thus, this ICME was not noted as an ICME that could have resulted from a 1 AU impact of the CME observed by HI, in this paper. Similarly, Braga et al. (2020) matched the 26 October CME (CME 12) to the ICME on 01 November, which we felt was better suited to the arrival of the 29 October CME; the 26 October CME was, we felt, more likely to be associated with the 30 October ICME from the ACE Caltech catalogue (though this event was not recorded in the Wind list), for the reasons given above.

**10 Summary**

The aim of this study is more to do with how CMEs are identified than with their detailed physics. HIGeoCAT is a catalogue of STEREO HI CMEs compiled by the HI Principal Investigator team and employing established, albeit simplistic, geometrical modelling techniques. The CDAW catalogue is a long-established catalogue of LASCO CMEs compiled by the LASCO team. CME identification and characterisation is at the heart of space weather forecasting and comparisons of these two CME lists can provide insights into (1) our ability to detect accurately and track potentially Earth-directed CMEs, (2) comparing observations made off the Sun-Earth line to those made near-Earth, and (3) our ability to identify and track CMEs from the corona into the heliosphere.

The need to select a period with off Sun-Earth line observations from the STEREO spacecraft, with the Sun in a reasonably active state, led to the selection of the calendar year 2011. Of the 116 HIGeoCAT HI-A CMEs identified in 2011, 35 appear to be associated with halo or partial halo CMEs listed in the CDAW catalogue. This suggests a reasonable consistency between the Earth-directed CMEs detected from both on and off Sun-Earth line vantage points which should be explored more in a further study. An example event of this type, that of 14 February 2011, is described in Section 5.

The key findings of this study can be listed as the following:

- A significant subset of HIGeoCAT HI-A CMEs that were deemed to be likely to be Earth impacting were identified, based on their estimated propagation direction from SSEF analysis. Of the 28 HIGeoCAT HI-A CMEs for which this was the case, 15 CMEs were identified that were potentially Earth directed for which no corresponding halo or partial halo CME was identified for the LASCO data. It is the analysis of these events that lies at the heart of this paper.

- In a year where there are only 32 and 19 ICME arrivals at L1 reported in the Caltech ACE and NASA Wind catalogues, respectively, such a large set of discrepant CMEs that are potentially Earth-directed, where there is ambiguity between observations on and off the Sun-Earth line, is of concern.

- We find that for 9 of the 15 anomalous Earth-directed CMEs, there is consistency with an ICME arrival at Earth, despite the lack of a consistent CDAW halo/partial halo CME entry. The success in identifying these as Earth-directed CMEs can be attributed to the use of an off Sun-Earth line vantage point, combined with the derivation of a propagation direction from analysis of the HI data.

- For these 9 events, the time difference between the estimated 1 AU arrival time (using the SSEF30 technique applied to the HI-A data) compared to the actual arrival time (defined by the first recorded shock or ICME onset of the ICME nearest in time, as defined in Table 1) ranges from 2 hours 7 minutes to 35 hours 37 minutes. The average absolute difference is 11.4 hours. However, the time difference figures are dominated by the two outlier differences for events 9 and 11. The average absolute difference between the estimated 1 AU arrival time and the actual ICME onset for the remaining events was 5.2 hours. For three of the events, the SSEF30-estimated arrival preceded the actual ICME arrival; for the remaining 6, the actual ICME arrival preceded the SSEF30-estimated arrival.

• For the remaining 6 of the 15 inconsistent CMEs, the HIGeoCAT HI-A catalogue provided evidence of potentially Earth-directed CMEs, but no ICME arrival was recorded. In each case, there was no corresponding CDAW halo/partial CME. These could be near-miss events and, from a forecasting point of view, would be of interest. However, without evidence for the impact at Earth, we should understand that, in some cases, there might have been errors in the consistent tracking of the CME front or cases where the SEFF30 assumptions are not well suited to the CME.

• Of the 9 HIGeoCAT HI-A CMEs with consistent ICME arrival observations at 1 AU, 4 showed evidence of preceding CME remnants, one showed an adjacent CME structure, and another was followed by another CME. Such events could potentially compromise head-on CME identification from L1 due to the superposition of CME structures, compounded by the inherent faint nature of halo CMEs. These observations underline the value of observations in the corona and heliosphere from off the Sun-Earth line in conjunction with coronagraph observations from L1.

This work illustrates that, for credible space weather impact prediction, both on and off Sun-Earth line observations, including heliospheric imaging, are essential for a mature space weather monitoring capability.

**Data Availability Statement**

This work makes use of catalogues from three sources and the authors acknowledge the support of providers of these catalogues. This includes the HELCATS catalogues (https://www.helcats-fp7.eu/), and, in particular the STEREO HI instrument CME catalogues, HICAT (https://www.helcats-fp7.eu/catalogues/wp2_cat.html), HIGeoCAT (https://www.helcats-fp7.eu/catalogues/wp3_cat.html) and HIJoinCAT (https://www.helcats-fp7.eu/catalogues/wp2_joincat.html). The HICAT and HIGeoCAT catalogues are generated and maintained by the STEREO HI instrument team at the STFC Rutherford Appleton Laboratory. The study also makes use of the CDAW SOHO LASCO instrument CME catalogue (https://cdaw.gsfc.nasa.gov/CME_list/), which is generated and maintained at the CDAW Data Center by NASA and The Catholic University of America in cooperation with the Naval Research Laboratory. We also acknowledge the STEREO COR-2 catalogue (http://solar.jhuapl.edu/Data-Products/COR-CME-Catalog.php), which is generated and maintained at Johns Hopkins University Applied Physics Laboratory, in collaboration with the Naval Research Laboratory and the NASA Goddard Space Flight Center, and is supported by NASA.

**Acknowledgments**

The HI instruments aboard STEREO were developed by a consortium that comprised the University of Birmingham (UK), the Rutherford Appleton Laboratory (UK), Centre Spatial de Liège (CSL, Belgium), and the Naval Research Laboratory (NRL, USA). The STEREO/SECCHI project, of which HI is a part, is an international consortium led by NRL. The LASCO instrument was constructed by a consortium comprising NRL, the University of Birmingham, the Max Planck Institut fur Aeronomie (MPAe, Germany) and the Laboratoire d'Astronomie Spatiale (LAS, France). We acknowledge the EU FP7 HELCATS (Heliospheric Cataloguing, Analysis and Techniques Service) project (www.helcats-fp7.eu/),

which was supported by the European Union FP7–SPACE–2013–1 programme (project #606692). We also acknowledge the LASCO CME catalogue generated and maintained at the CDAW Data Center by NASA and The Catholic University of America in cooperation with NRL, and the COR-2 catalogue incorporated into the Dual-Viewpoint Civil Catalog from the SECCHI/COR-2 Telescopes of the Applied Physics Laboratory of the Johns Hopkins University. RAH, JAD and DB acknowledge support via the RAL Space In House Research programme funded by the Science and Technology Facilities Council of UK Research and Innovation. CM thanks the Austrian Science Fund (FWF): P31659-N27, funded by the European Union (ERC, HELIO4CAST, 101042188). Views and opinions expressed are, however, those of the author(s) only and do not necessarily reflect those of the European Union or the European Research Council Executive Agency. Neither the European Union nor the granting authority can be held responsible for them.

Table 1. Key parameters of the HIGeoCAT (SSEF30) and CDAW-listed CMEs for which the projected HI-A CME source region lies within the +/- $20^{o}$ Target Zone centred on the centre of the solar disc as viewed from Earth, and for which there is no associated CDAW partial halo or halo CME. Correspondong HI-B CMEs are identified using HIJoinCAT. CDAW events are listed that have reported projected onsets within +/- 2 hours of the HI-A-derived CME onset time (at 1 Rs). The projected 1 AU arrival of each HI CME assumes the HIGeoCAT constant speed is maintained out to Earth. ACE and Wind ICME arrivals at L1 are noted if the SSEF30 projected CME 1 AU arrival time (estimated from the HI-A data) is within 48 hours of the ICME shock, or onset and end times.

| | HELCATS Event ID | HIGeoCAT (STEREO HI) CME | | | | | | | CDAW (LASCO) CME | | | ACE ICME | | Wind ICME | |
|---|---|---|---|---|---|---|---|---|---|---|---|---|---|---|---|
| | | Time of first observation in HI-1 field of view (UT) | SSEF30: Source longitude HEEQ (HEE) (degrees) | SSEF30: Source latitude HEEQ (HEE) (degrees) | SSEF30: CME onset (Sun Centre) (UT) | SSEF30: Projected onset at 1Rs (UT) | SSEF30 model: CME radial speed (km/s) | SSEF30 model: Arrival at 1 AU (UT) | Time of first observation in LASCO field of view | Back-projected CME onset (at limb, plane of sky) | CME speed (km/s) | Time of shock (onset and end times of ICME) | ICME speed at 1 AU | Time of ICME onset (magnetic obstacle onset and end time of ICME) | ICME Speed at 1 AU |
| 1. | HCME_A__2011 0124_01 | 24 Jan 10:09 | 12 (12) | -6 (0) | 24 Jan 03:35 | 4:06 | 380 | 28 Jan 17:14 | 24 Jan 02:00 | 2:15 | 258 | No event reported: 24 Jan to 4 Feb | n/a | No event reported: 24 Jan to 18 Feb | n/a |
| | HCME_B__2011 0124_01 | 24 Jan 12:09 | 15 (14) | -1 (5) | 24 Jan 01:47 | 2:17 | 388 | 28 Jan 13:13 | | | | | | | |
| 2. | HCME_A__2011 0314_01 | 14 Mar 08:49 | 7 (7) | 0 (7) | 14 Mar 02:44 | 3:04 | 577 | 17 Mar 02:57 | 14 Mar 04:24 | 3:48 | 415 | No event reported: 6 to 29 Mar | n/a | No event reported: 18 Feb to 29 Mar | n/a |
| | No HIJoinCAT HI-B CME listed | n/a | n/a | n/a | n/a | n/a | n/a | n/a | | | | | | | |
| 3. | HCME_A__2011 0320_01 | 20 Mar 02:09 | -7 (-7) | -20 (-13) | 19 Mar 17:24 | 18:03 | 299 | 25 Mar 12:45 | 19 Mar 16:48 | 16:50 | 602 | No event reported: 6 to 29 Mar | n/a | No event reported: 18 Feb to 29 Mar | n/a |
| | HCME_B__2011 0320_01 | 20 Mar 02:09 | -14 (-14) | -29 (-23) | 19 Mar 19:43 | 20:03 | 594 | 22 Mar 17:52 | | | | | | | |
| 4. | HCME_A__2011 0503_01 | 03 May 22:09 | -4 (-3) | 9 (14) | 03 May 17:29 | 17:47 | 648 | 6 May 09:47 | 03 May 16:51<br>03 May 17:43 | 16:00<br>16:45 | 658<br>346 | No event reported: 29 Mar to 28 May | n/a | No event reported: 29 Apr to 28 May | |
| | HCME_B__2011 0503_02 | 03 May 21:29 | -16 (-14) | 18 (24) | 03 May 16:02 | 16:26 | 495 | 7 May 04:13 | | | | | | | |
| 5. | HCME_A__2011 0525_01 | 25 May 08:09 | -17 (-18) | -11 (-8) | 25 May 03:15 | 3:37 | 545 | 28 May 07:42 | 25 May 05:24 | 4:30 | 276 | | 510 | | 519 |
| | HCME_B__2011 0525_01 | 25 May 08:49 | 0 (-1) | -14 (-13) | 25 May 00:53 | 1:16 | 498 | 28 May 12:33 | | | | 28 May 01:00 (28 May 05:00 to 21:00) | | 28 May 00:14 (28 May 05:31 to 22:47) | |
| 6. | HCME_A__2011 0604_01 | 04 Jun 02:09 | 2 (0) | -20 (-20) | 03 Jun 20:17 | 20:46 | 412 | 8 Jun 01:25 | 03 Jun 21:24<br>03 Jun 22:00 | 19:45<br>20:40 | 303<br>239 | No event reported: 4 to 17 Jun | n/a | No event reported: 4 to 17 Jun | n/a |
| | HCME_B__2011 0604_01 | 04 Jun 03:29 | -1 (-3) | -19 (-19) | 03 Jun 21:02 | 21:28 | 448 | 7 Jun 18:02 | | | | | | | |
| 7. | HCME_A__2011 0606_01 | 06 Jun 14:09 | 7 (8) | 10 (9) | 06 Jun 08:47 | 9:17 | 382 | 10 Jun 21:52 | 06 Jun 07:30<br>06 Jun 10:27 | 7:45<br>9:06 | 582<br>249 | No event reported: 4 to 17 Jun | n/a | No event reported: 4 to 17 Jun | n/a |
| | No HIJoinCAT HI-B CME listed | n/a | n/a | n/a | n/a | n/a | n/a | n/a | | | | | | | |
| 8. | HCME_A__2011 0623_01 | 23 Jun 20:09 | -10 (-8) | 17 (17) | 23 Jun 13:24 | 14:08 | 266 | 30 Jun 02:02 | 23 Jun 13:25 | 12:08 | 347 | No event reported: 17 Jun to 6 Jul | n/a | 30 Jun 13:26 (30 Jun 18:43-1 Jul 05:16) | 329 |
| | HCME_B__2011 0623_01 | 23 Jun 19:29 | -21 (-21) | -2 (-2) | 23 Jun 09:51 | 10:22 | 381 | 27 Jun 23:13 | | | | | | | |
| 9. | HCME_A__2011 0703_01 | 03 Jul 02:49 | -5 (-6) | -11 (-14) | 02 Jul 19:50 | 20:07 | 713 | 5 Jul 06:16 | 02 Jul 20:36 | 19:30 | 310 | 6 Jul 02:58 (6 Jul 17:00 to 7 Jul 12:00) | 360 | 3Jul 19:12 (3 Jul 19:12 to 4 Jul 19:12) | 392 |
| | No HIJoinCAT HI-B CME listed | n/a | n/a | n/a | n/a | n/a | n/a | n/a | | | | | | | |
| 10. | HCME_A__2011 0711_01 | 11 Jul 16:49 | -8 (-9) | -1 (-5) | 11 Jul 08:47 | 9:11 | 494 | 14 Jul 21:07 | 11 Jul 10:00<br>11 Jul 12:00 | 8:45<br>10:50 | 268<br>266 | 14 Jul 12:00 (15 Jul 04:00 to 16 Jul 15:00) | 410 | No event reported: 3 Jul to 17 Sep | n/a |
| | HCME_B__2011 0711_01 | 11 Jul 17:29 | -22 (-23) | -2 (-4) | 11 Jul 04:25 | 4:57 | 361 | 15 Jul 23:50 | | | | | | | |
| 11. | HCME_A__2011 1005_01 | 05 Oct 01:29 | 13 (13) | 0 (-5) | 04 Oct 19:25 | 19:50 | 458 | 8 Oct 14:24 | 04 Oct 19:48<br>04 Oct 21:24 | 19:10<br>20:20 | 416<br>501 | 6 Oct 12:00 (6 Oct 12:00 to 23:00) and 7 Oct 02:47 (7 Oct 02:00 to 17:00) | 370 and 380 | 5 Oct 06:46 (5 Oct 09:56 to 7 Oct 02:06) | 399 |
| | No HIJoinCAT HI-B event listed | n/a | n/a | n/a | n/a | n/a | n/a | n/a | | | | | | | |
| 12. | HCME_A__2011 1026_01 | 26 Oct 16:49 | -4 (-4) | 6 (2) | 26 Oct 11:06 | 11:32 | 449 | 30 Oct 07:54 | 26 Oct 11:48<br>26 Oct 13:25 | 11:00<br>11:50 | 324<br>262 | 30 Oct 10:01 (31 Oct 01:00 to 15:00) | 449 and 380 | No event reported: 24 Oct to 1 Nov | n/a |
| | HCME_B__2011 1026_01 | 26 Oct 20:09 | -87 (-87) | 6 (1) | 25 Oct 23:44 | 00:10 | 451 | 29 Oct 20:33 | | | | | | | |
| 13. | HCME_A__2011 1029_01 | 29 Oct 02:49 | 10 (8) | 15 (12) | 28 Oct 17:31 | 17:57 | 447 | 1 Nov 14:44 | 28 Oct 19:00 | 17:36 | 225 | 30 Oct 10:01 (31 Oct 01:00 to 15:00) or 1 Nov 09:07 (2 Nov 01:00 to 3 Nov 04:00) | 449 and 380 | 1 Nov 08:09 (1 Nov 08:09 to 22:04) or 2 Nov 00:21 (2 Nov 03:35 to 3 Nov 04:33) | 412 and 371 |
| | HCME_B__2011 1029_01 | 29 Oct 02:10 | 13 (10) | 21 (18) | 28 Oct 16:58 | 17:20 | 526 | 1 Nov 00:10 | | | | | | | |
| 14. | HCME_A__2011 1101_01 | 01 Nov 12:09 | -9 (-10) | 20 (16) | 1 Nov 05:59 | 6:25 | 450 | 5 Nov 02:35 | 01 Nov 08:46 | 08:00 | 400 | | n/a | | 303 |
| | No HIJoinCAT HI-B event listed | n/a | n/a | n/a | n/a | n/a | n/a | n/a | | | | No event reported: 1 to 7 Nov. | | 4 Nov 20:27 (4 Nov 21:17 to 5 Nov 20:09) | |
| 15. | HCME_A__2011 1129_01 | 29 Nov 03:29 | 1 (0) | 11 (11) | 29 Nov 00:05 | 0:32 | 437 | 2 Dec 23:21 | No event | n/a | n/a | 2 Dec 17:17 (2 Dec 18:00 to 3 Dec 07:00) | 400 | No event reported: 28 Nov to 21 Jan | n/a |
| | HCME_B__2011 1129_01 | 29 Nov 04:50 | -11 (-14) | 24 (22) | 28 Nov 20:16 | 20:47 | 377 | 3 Dec 10:47 | | | | | | | |

Figure 1. HI-1A (left) and HI-1B (middle) difference images of the 14 February 2011 CME, both from 17:29 UT on 15 February 2011. As with all of the HI images presented in this paper, we show the full $20^{o}$ x $20^{o}$ field of view of each HI-1 camera. The Sun is $4^{o}$ off the right/left hand side of each HI-1A/HI-1B image and the ecliptic plane runs approximately across the horizontal centre-line of each image. The right-hand image shows the CDAW halo CME of 15 February.

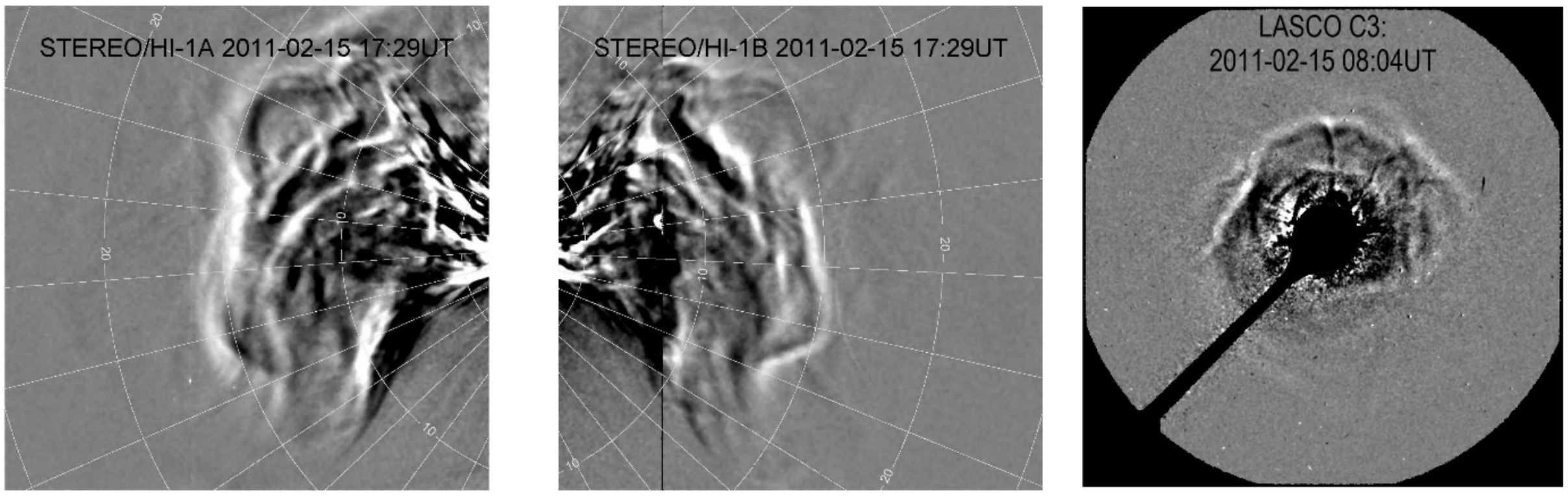

Figure 2. Selected difference images from HI-1A (top) and HI-1B (middle) of the 11 July 2011 HI CME (CME 10 in Table 1). Times are given in each frame. The bottom image shows the same CME in a STEREO-A COR-2 (COR-2A) coronagraph difference image from 14:09 UT on 11 July.

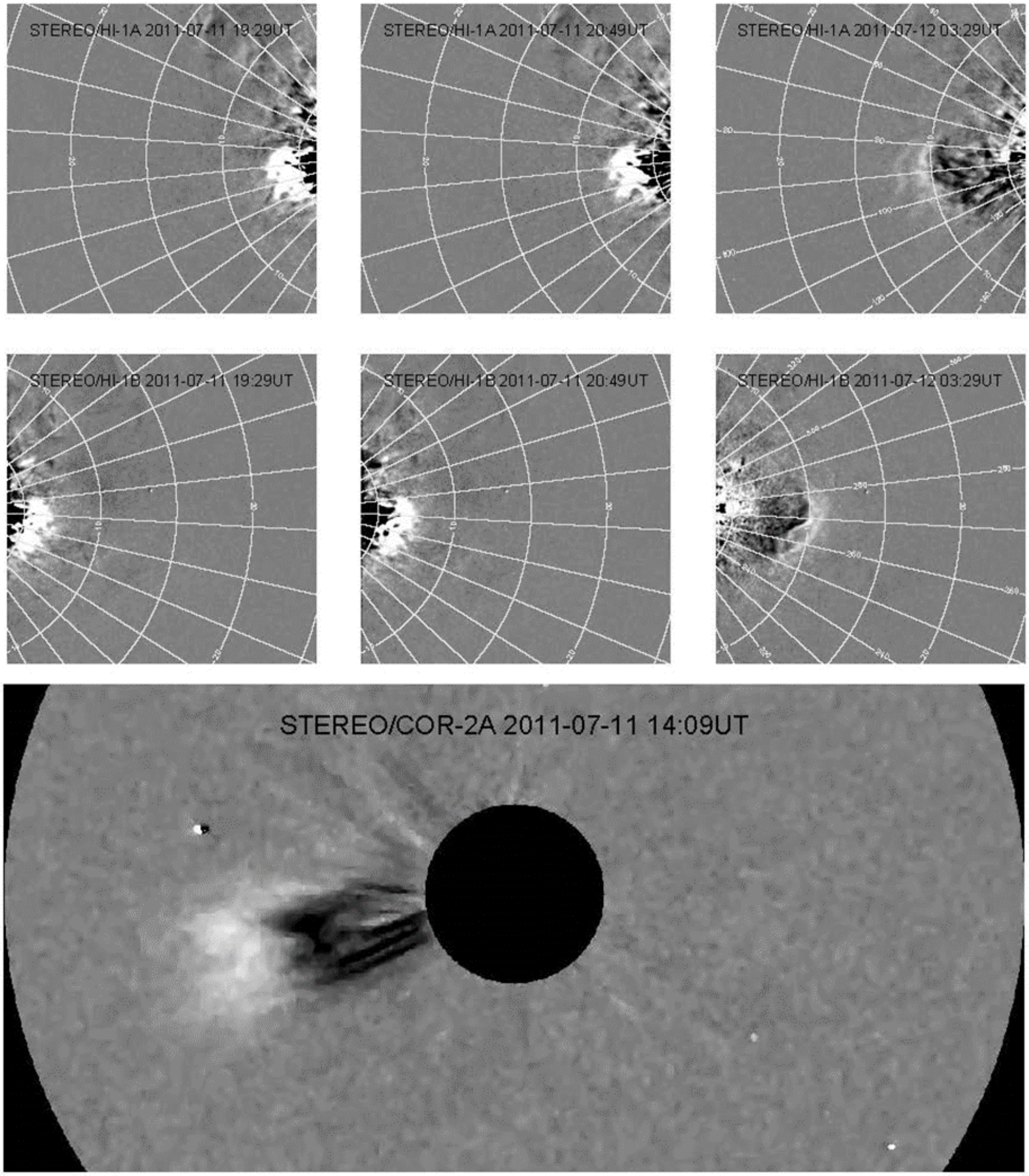

Figure 3.  Selected difference images from HI-1A (top) and HI-1B (middle) of the 25 May HI CME (CME 5) in the same format as Figure 1. The bottom image shows the same CME in the COR-2A data, at 05:39 UT.

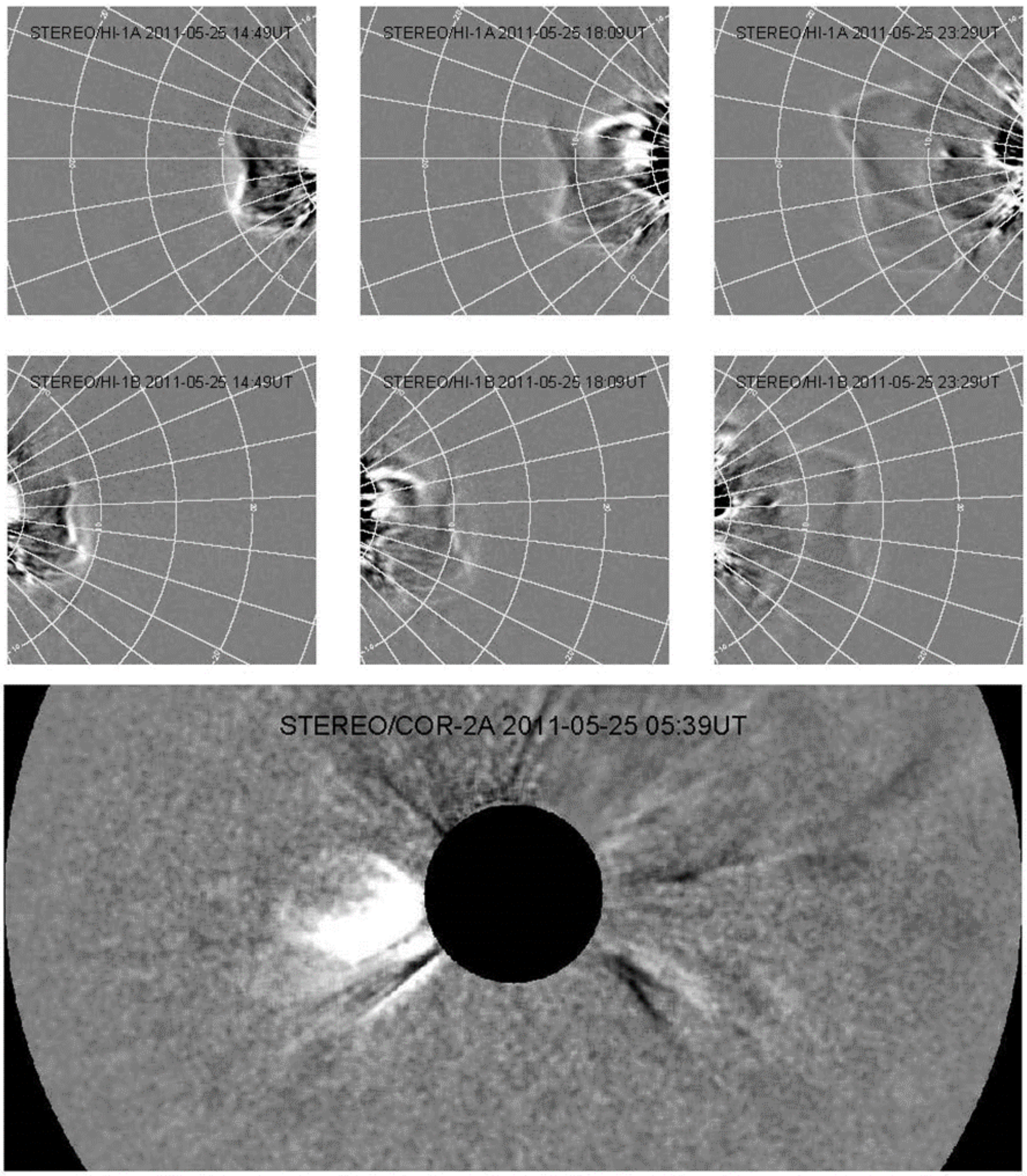

Figure 4. A LASCO C2 difference image showing the CME off the north-west limb, with projected onset at 04:30 UT on 25 of May.

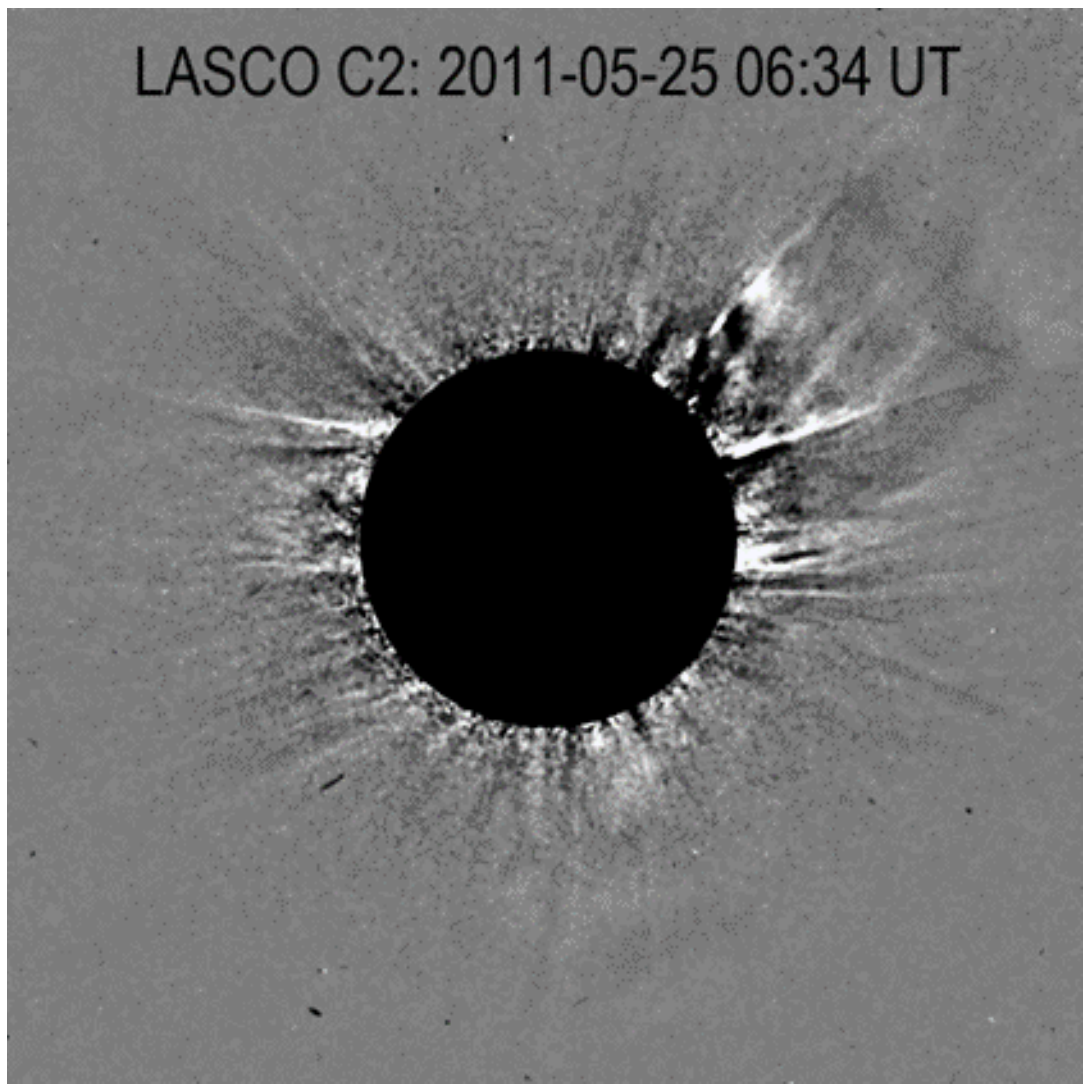

Figure 5. HI-1A (left), HI-1B (middle) and COR-2A (right) difference images for the events first observed in HI data on 3 July (CME 9; top), 26 October (CME 12; middle) and 29 November (CME 13; bottom), 2011.

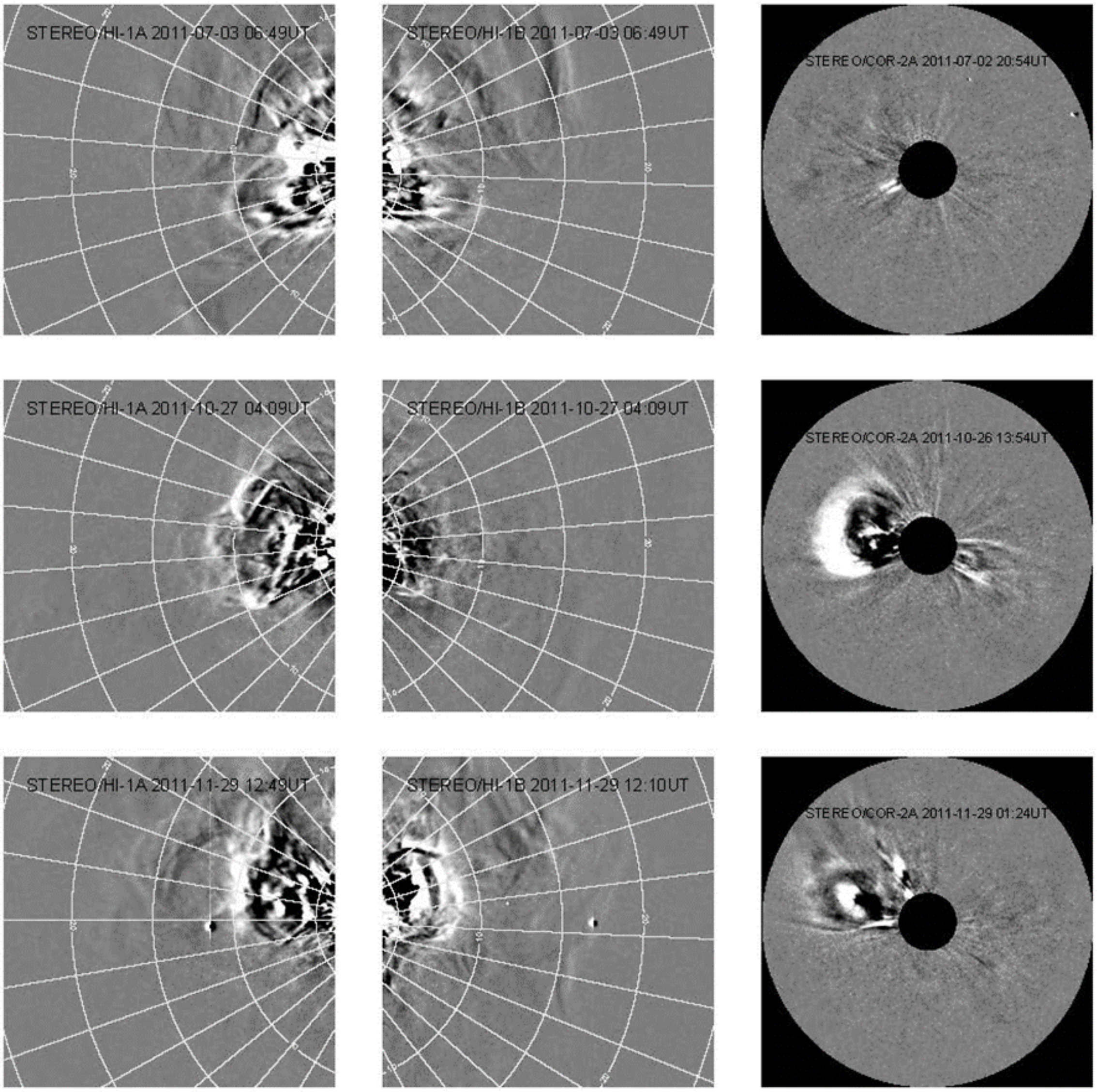

Figure 6. HI-1A  (top left) and HI-1B (top right) difference images of the 24 January CME (CME 1). A COR-2A difference image at 09:24 UT is included (middle), as is a LASCO C3 difference image at 11:16 UT (bottom left) and an SDO AIA image at 01:00 UT (bottom right).

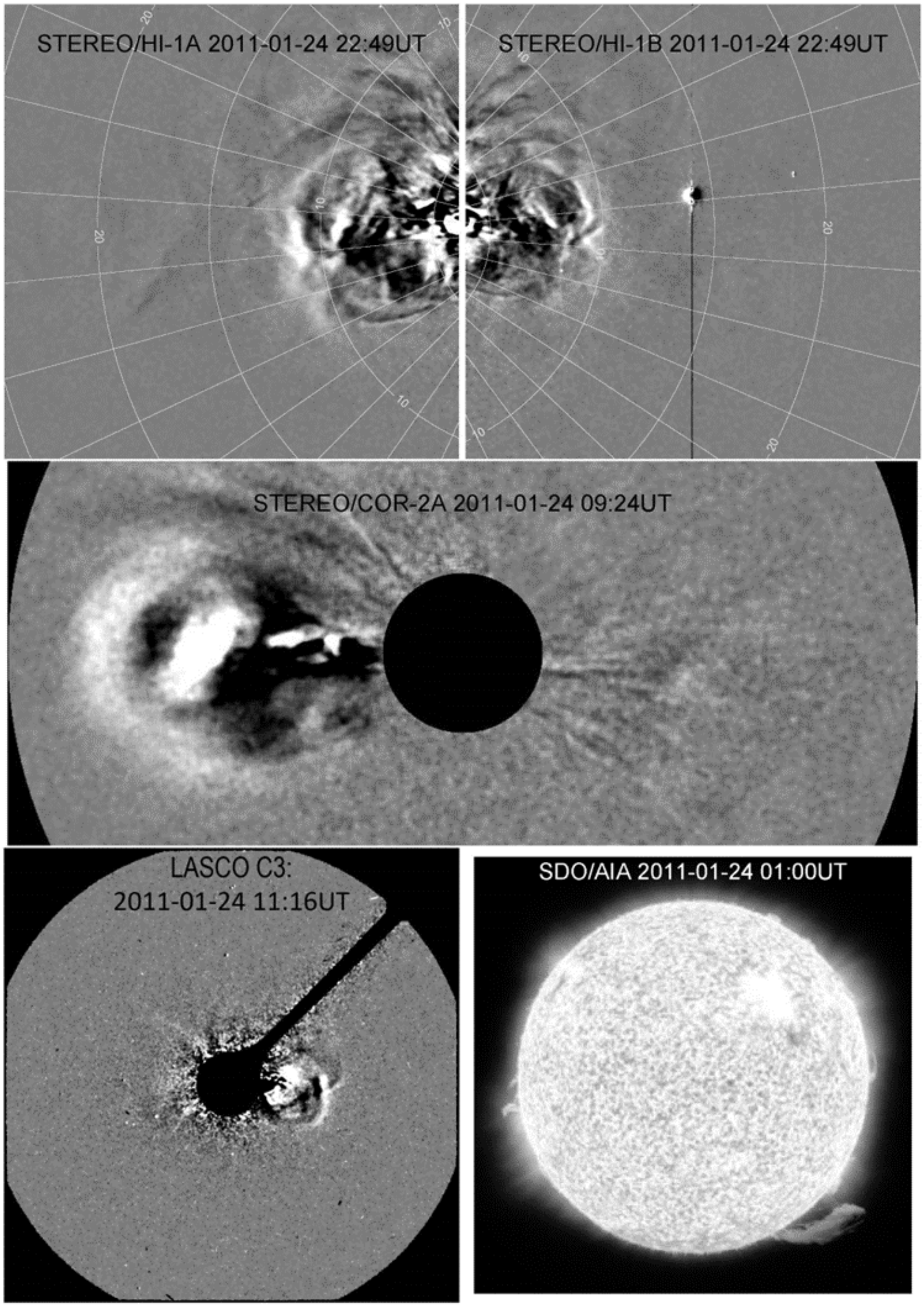